\documentclass[%
 twocolumn,
 superscriptaddress,
 amsmath,amssymb,
 aps,
prb,
floatfix,10pt
]{revtex4-2}

\usepackage[utf8]{inputenc}
\usepackage{graphicx}% Include figure files
\usepackage{dcolumn}% Align table columns on decimal point
\usepackage{bm}% bold math
\usepackage{amsmath}
\usepackage{wasysym}
\usepackage{xcolor}
\usepackage{hyperref}
\usepackage{url}
\usepackage[normalem]{ulem}
\usepackage[dvipsnames]{xcolor}
\usepackage[percent]{overpic}

\begin{document}

\preprint{APS/123-QED}

\title{Ab Initio Investigation of Pressure Effects in the Spin-Liquid Candidate Y-Kapellasite}% Force line breaks with \\

\author{Federico Abbruciati}
 \email{federico.abbruciati@esrf.fr}
 \affiliation{European Synchrotron Radiation Facility, 71 Avenue des Martyrs, F-38043 Grenoble, France}
 \affiliation{Institute for Quantum Materials and Technologies, Karlsruhe Institute of Technology, Kaiserstr. 12, 76131
Karlsruhe, Germany.}

\author{Aleksandar Razpopov}
\email{razpopov@itp.uni-frankfurt.de}
\affiliation{Institut f\"ur Theoretische Physik, Goethe-Universitat Frankfurt, 60438 Frankfurt am Main, Germany}

\author{João Elias F. S. Rodrigues}
\affiliation{European Synchrotron Radiation Facility, 71 Avenue des Martyrs, F-38043 Grenoble, France}

\author{Gaston Garbarino}
\affiliation{European Synchrotron Radiation Facility, 71 Avenue des Martyrs, F-38043 Grenoble, France}

\author{Matthieu Le Tacon}
\affiliation{Institute for Quantum Materials and Technologies, Karlsruhe Institute of Technology, Kaiserstr. 12, 76131
Karlsruhe, Germany.}

\author{Roser Valent\'\i}
\email{valenti@itp.uni-frankfurt.de}
\affiliation{Institut f\"ur Theoretische Physik, Goethe-Universitat Frankfurt, 60438 Frankfurt am Main, Germany}

\author{Pascal Puphal}
\affiliation{Max Planck Institute for Solid State Research, Heisenbergstraße 1, D-70569 Stuttgart, Germany}
 
\author{Björn Wehinger}%
 \email{bjorn.wehinger@esrf.fr}
 \affiliation{European Synchrotron Radiation Facility, 71 Avenue des Martyrs, F-38043 Grenoble, France}

\date{\today}% It is always \today, today,
             %  but any date may be explicitly specified

\begin{abstract}

Motivated by recent experiments showing pressure-induced suppression of magnetic order and the emergence of a dynamical ground state in the anisotropic kagome antiferromagnet Y-kapellasite Y$_3$Cu$_9$(OH)$_{19}$Cl${_8}$, we perform ab initio density functional theory (DFT) calculations to investigate the evolution of magnetic exchange interactions under hydrostatic pressure. We show that pressure efficiently tunes the magnetic Hamiltonian by altering the Cu–O–Cu bond geometry, thereby driving the system towards a spin-liquid regime. This evolution is governed by a nonlinear dependence of the dominant exchange coupling on the Cu–O–Cu bond angle. We further examine the influence of hydrogen positions and find that both the O–H bond length and the hydrogen out-of-plane angle strongly affect the magnetic interactions. Our results provide a microscopic explanation for the experimentally observed pressure-induced enhancement of frustration and highlight the key role of hydrogen geometry.

%Magnetic frustration in quantum materials provides a fertile ground for exotic quantum states, promoting strong quantum fluctuations, fractionalized excitations, emergent gauge phenomena, and suppression of long-range order at zero temperature. In this context, proper tuning of magnetic interactions is a key element to access the possible quantum states. Here we study the pressure behavior of the frustrated anisotropic kagome system Y-kapellasite Y$_3$Cu$_9$(OH)$_{19}$Cl${_8}$. We perform \textit{ab initio} density functional theory (DFT) total energy mapping calculations to extract magnetic exchange couplings, showing how hydrostatic pressure acts as an efficient tool to modulate magnetism and drive the system towards a spin liquid state. We find that pressure-induced changes in the Cu–O–Cu bond geometry drive the evolution of J, which follows a nonlinear angle dependence consistent with previous theoretical studies. To address the low X-ray scattering cross section of hydrogen, we further investigate the dependence of the magnetic couplings on the O–H bond length and the hydrogen out-of-plane angle $\tau$. Our results highlight the crucial role of hydrogen positioning in accurately determining the magnetic interactions. \rv{relation with experiments?}

\end{abstract}

%\keywords{Suggested keywords}%Use showkeys class option if keyword
                              %display desired
\maketitle

\section{Introduction}

Interacting spins in quantum materials lead to very rich and diverse phase diagrams, often exhibiting unique and intriguing properties such as quantum spin liquids, long-range entangled phases, and fractionalized excitations~\cite{balents2010spin,savary2016quantum,zhou2017quantum,rousochatzakis2017classical,knolle2019field,broholm2020quantum}. Being able to tune and control magnetic interactions is therefore essential to access and explore the resulting states of matter. Pressure has proven to be a valuable knob to modulate magnetic properties in low-dimensional systems where spin interactions are largely governed by the underlying crystal structure \cite{thede2014pressure,ruegg2004pressure,biesner2019pressure,malavi2020pressure,wehinger2018giant}.
In this context, frustrated magnetism serves as an exciting platform for exploration.
Pressure-induced lattice distortions tune magnetic interactions, altering frustration, ground-state degeneracy, and quantum fluctuations. 
 
 Since Anderson’s proposal of the resonating valence bond state \cite{anderson1973resonating}, 
 %and Kitaev’s exactly solvable model \cite{kitaev2006anyons}, 
 quantum spin liquids have been a major focus of condensed matter physics and quantum information science. This has motivated extensive studies of geometrically frustrated magnets, particularly S=1/2 triangular-lattice~\cite{tocchio2013spin,kaneko2014gapless,iqbal2016spin,zhu2015spin,scheie,riedl2022ingredients,bag2024evidence} and kagome systems~\cite{mendels2016quantum,norman2016colloquium}.
The attention on kagome lattices was brought by herbertsmithite ZnCu$_3$OH$_6$Cl$_2$~\cite{mendels2010quantum}, a Cu$^{2+}$ $S=1/2$ kagome material, where no evidence of long-range magnetic order was found down to the lowest temperature. The material has been extensively studied both theoretically and experimentally \cite{han2012fractionalized,jeschke2013first,mendels2016quantum,norman2016colloquium,khuntia2020gapless}. However, its magnetic response is strongly affected by impurity spins arising from Cu$^{2+}$/Zn$^{2+}$ intersite disorder \cite{freedman2010site,kremer2025chemical}. While impurities can reveal emergent phenomena in quantum materials, they hinder the realization of an ideal $S=1/2$ kagome antiferromagnet. The polymorph kapellasite was proposed as a solution, as its layered structure suppresses Cu/Zn mixing and free-spin formation. Although no long-range magnetic order was observed down to 20 mK, subsequent studies revealed residual intersite disorder in this material as well~\cite{colman2008toward,colman2010comparisons,faak2012kapellasite}.

\begin{figure}
\centering
    \begin{overpic}[width=0.98\linewidth]{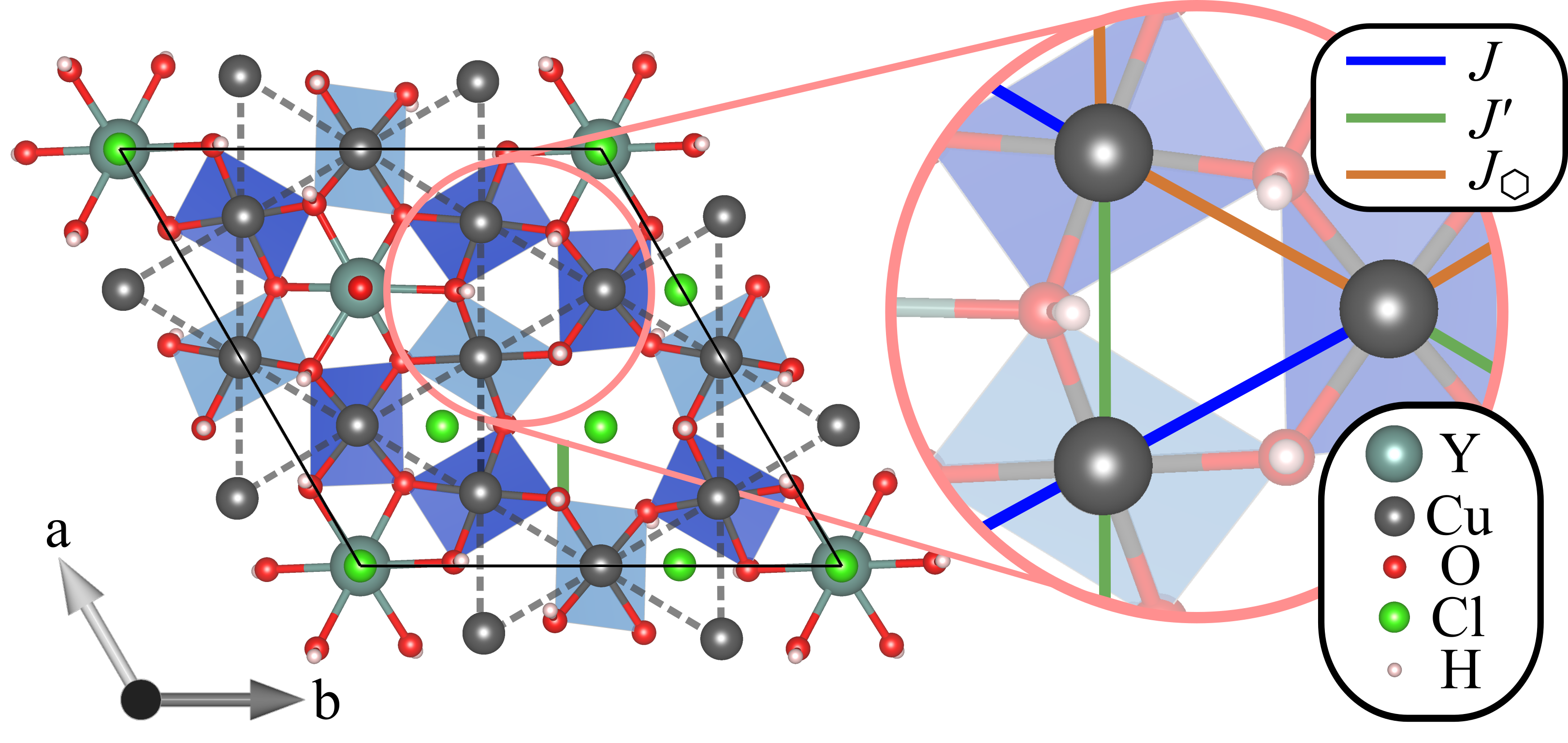}%
    \put(0,48){ \small (a)}
    \end{overpic}
    \par\vspace{0.7cm}
    \begin{overpic}[width=0.98\linewidth]{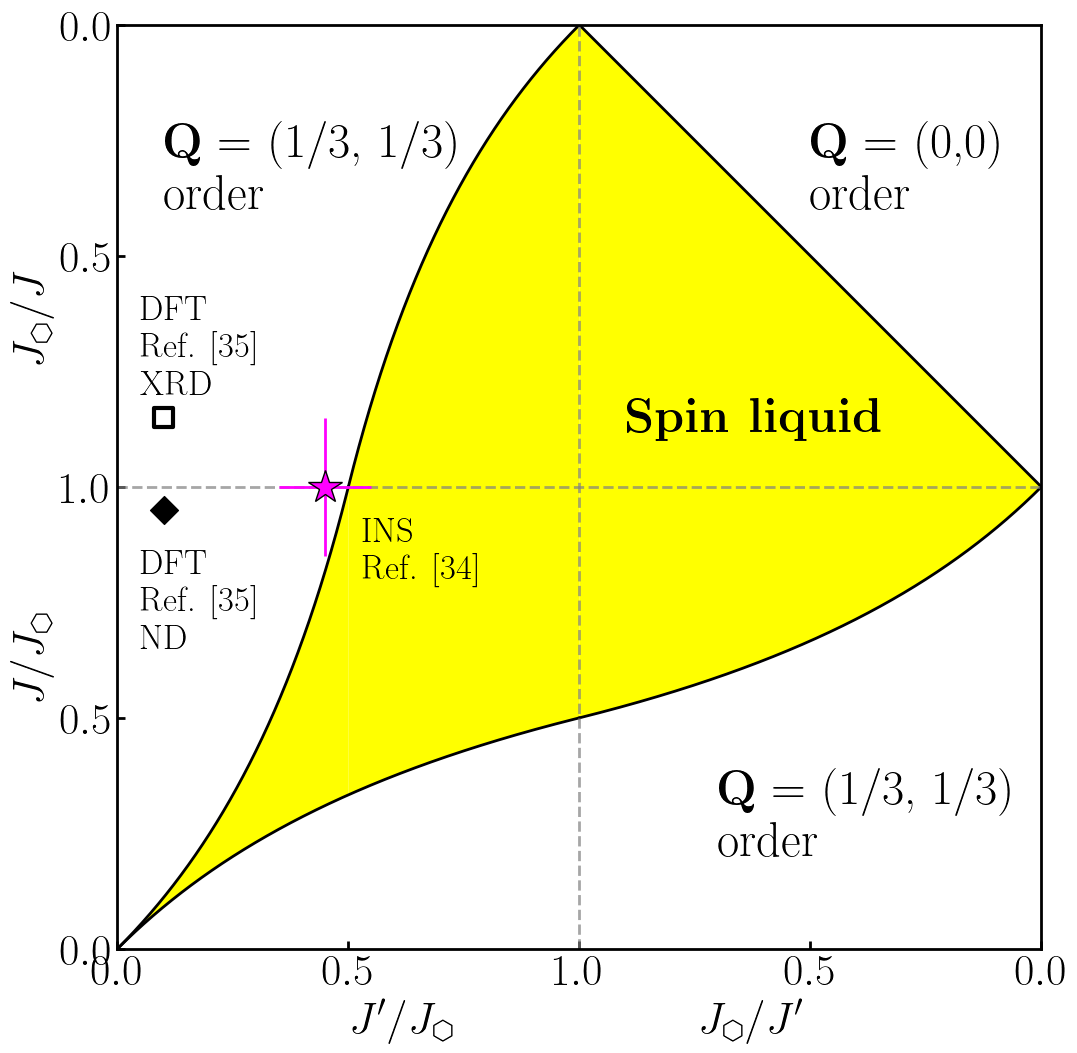}
    \put(0,100){ \small (b)}
    \end{overpic}
    \caption{(a) A top view of Y-kapellasite showing the kagome lattice in the ab plane with a zoom on the local geometry of the three main magnetic interactions $J'$, $J_{\varhexagon}$, and $J$. The two non-equivalent copper positions Cu1 and Cu2 are represented by the blue and light-blue environment. The superexchange coupling is mediated by the non magnetic hydroxide ion between the magnetic Cu. (b) Phase diagram of the classical $J_1-J_2-J_3$ (denoted $J'$, $J_{\varhexagon}$, and $J$ in this work) kagome  \cite{hering2022phase}, predicted by classical Monte Carlo simulations. The system features two coplanar states with ordering vectors \textbf{Q} = (1/3, 1/3, 0) and \textbf{Q} = (0, 0), and a spin liquid state according to the ratio of the first three nearest-neighbor couplings. Prior DFT on x-ray diffraction structure (XRD) and neutron diffraction structure (ND) and INS results \cite{chatterjee2023spin} are plotted on the phase diagram.}
    \label{fig:structure}
\end{figure}

These findings stimulated the exploration of related crystal families, leading to the growth of single crystals of Y-kapellasite, Y$_3$Cu$_9$(OH)$_{19}$Cl${_8}$, an antiferromagnetic quasi-two-dimensional distorted kagome system~\cite{puphal2017strong,barthelemy2019local,biesner2022multi,chatterjee2023spin}. A major advantage over previously reported compounds is that the substantial difference in atomic number between Y and Cu suppresses intersite mixing. This anisotropic version of a kagome lattice features two non-equivalent Cu positions and, as a result, three different nearest neighbor exchange couplings $J'$, $J_{\varhexagon}$ and $J$. They are highlighted with different colors, alongside its structure, in Fig. \ref{fig:structure}(a). Each Cu atom is connected to the neighboring one via an oxygen, with resulting Cu-O-Cu angles ranging between  110$^{\circ}$ and 118$^{\circ}$, different for each one of the three exchange paths. Experimental investigations such as nuclear magnetic resonance (NMR), muon spin relaxation ($\mu$SR), and inelastic neutron scattering (INS) measurements, as well as theoretical studies~\cite{chatterjee2023spin,hering2022phase}, suggest that the system exhibits a long-range magnetic order characterized by the ground-state wave vector $\mathbf{Q}=(1/3,1/3,0)$. Analytical arguments and numerical estimates predict a more intriguing phase diagram when the exchange parameters $J'$/$J_{\varhexagon}$ and $J$/$J_{\varhexagon}$ are varied, featuring a spin liquid (SL) phase depending on the ratio of the three nearest-neighbor couplings~\cite{hering2022phase} [Fig. \ref{fig:structure}(b)]. 

In Y-kapellasite, magnetic interactions are mediated by hydroxide ions through superexchange. According to the Goodenough–Kanamori–Anderson rules \cite{kanamori1959superexchange,goodenough1955theory,anderson1950anti},exchange interactions in such geometries are expected to depend sensitively on the Cu-O-Cu bond angles and to change rapidly near 90°, with larger bond angles generally favoring stronger antiferromagnetic coupling. This makes Y-kapellasite particularly susceptible to structural tuning. This was confirmed by recent experiments showing that uniaxial strain relieves frustration and enhances antiferromagnetic order, whereas hydrostatic pressure appears to enhance frustration and drive the system toward a fluctuating, spin-liquid-like ground state above 2.3~GPa~\cite{wang2023controlled,chatterjee2026emergence}. This interpretation was supported by the pressure evolution of the crystal structure and by the assumption of an approximately linear dependence of the exchange couplings on the Cu-O-Cu angles.
However, the microscopic evolution of the magnetic interactions under pressure remains to be established. In particular, the exchange couplings in hydroxide-based kagome materials may depend not only on the Cu-O-Cu angles, but also on the positions of the hydrogen atoms, which are difficult to determine accurately in the experiments and may change under pressure. Here, we use density functional theory calculations based on experimentally determined crystal structures to determine the pressure dependence of the magnetic exchange interactions in Y-kapellasite. We find that pressure drives the calculated exchange parameters toward the spin-liquid region of the theoretical phase diagram, consistent with recent experiments. At the same time, we show that the dominant exchange interaction depends nonlinearly on the relevant structural parameters and that the calculated couplings are highly sensitive to the hydrogen positions. These results provide a microscopic basis for understanding pressure-tuned frustration in Y-kapellasite and highlight the crucial role of the local hydroxide geometry in determining its magnetic ground state.

\section{Methods} 
The experimentally determined structures from Ref. \cite{chatterjee2026emergence} were used as a starting point for DFT calculations.
For each structure we calculated the isotropic magnetic exchange couplings $J$ by applying total energy mapping analysis (TEMA) via spin-polarized DFT calculations \cite{jeschke2013first,glasbrenner2015effect,lowenergyHam,hering2022phase,razpopov2023j}. This method consists of a two step process, where first we calculated the total energy of different magnetic configurations via DFT and then in the second step, we mapped them to an effective Heisenberg Hamiltonian. 
Spin-polarized DFT energies were computed within the full potential localized orbital (FPLO) framework \cite{fplo1,fplo2}, employing the generalized gradient-approximation (GGA) as the exchange-correlation functional.  
Calculations were performed on the P1 unit cell with a minimum of 18 different magnetic configurations on a well-converged $k$-mesh with converging criteria of $10^{-8}$ Hartree on the energy and $10^{-6} \text{ e}^{-}/\text{\AA}^{3}$ on the density. 
The fractional occupancy of the H4 site (1/6) was treated within the FPLO virtual crystal approximation.
To account for the strongly localized Cu 3d electrons we use the GGA+U correction, applying the 'atomic limit' method~\cite{czyzyk1994local} and gross population projection, as implemented in FPLO. The Coulomb repulsion U was set within the range of 4 eV and 10 eV and the Hund's coupling $J_H \approx 1$ eV\cite{sugano2012multiplets,pavarini2011lda+}.
%\ar{Is it not exactly $J_H = 1$ eV? \fa{Jhund=((F2+F4)/14 and F2=8 F4=5 so that the realistic ratio is F4/F2=8/5 is kept for copper and Jh=13/14 approx 1)}}.
%The Slater integrals $F_2$ and $F_4$ were chosen to satisfy the common ratio $F_4/F_2 \approx 5/8$ for realistic 3d orbitals, U value was set to $F_0$ $J_H \approx 1$  . %
The magnetic exchange couplings $J_i$ were extracted via a least square fitting (see Sec. \ref{sec:pressure_dep} for more details).
%\ar{I do not think we need the errors statement in the methods section.}
%The errors indicate the uncertainty of the Hamiltonian parameters calculated for a given crystal structure; however, uncertainties in the crystal structure itself and their impact on the exchange couplings are not included. 
%For more details on the Hamiltonian describing the system, see Sec. \ref{sec:pressure_dep}.
The U value in the GGA+U calculation was varied until the appropriate set of couplings could reproduce the experimental Curie temperature of $\theta_{CW} = - 100$ K \cite{puphal2017strong}. The relation between $\theta_{CW}$ and the exchange couplings was evaluated within a mean-field framework \cite{suzukimean}.

\section{Results}
\subsection{Temperature and pressure dependence of the magnetic exchange couplings}
\label{sec:pressure_dep}

\begin{figure}
\centering

    \begin{overpic}[width=0.99\linewidth]{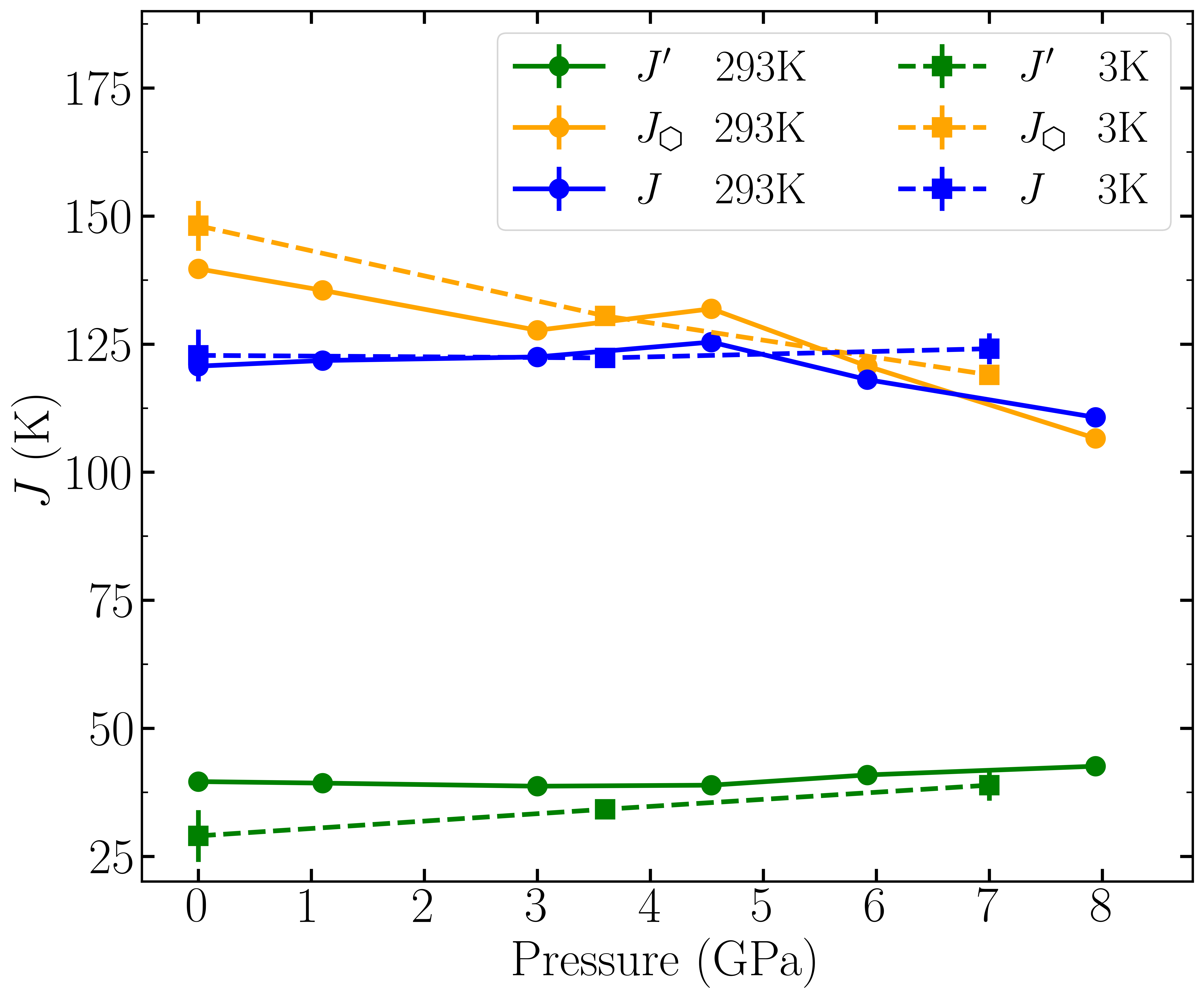}%
    \put(0,85){ \small (a)}
    \end{overpic}
    \par\vspace{0.5cm}
    \begin{overpic}[width=0.99\linewidth]{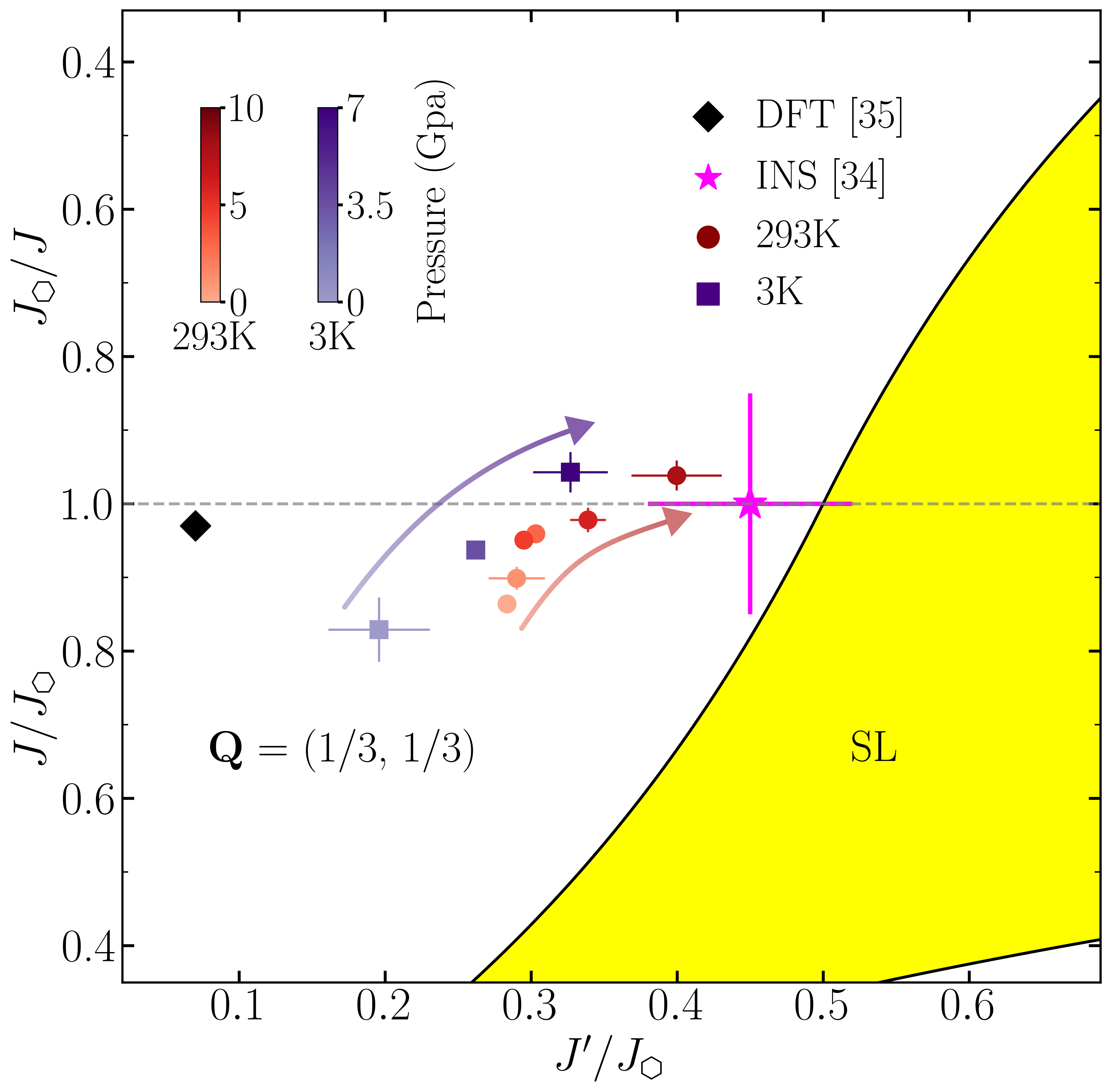}
    \put(0,100){ \small (b)}
    \end{overpic}

    \caption{(a) Pressure evolution of the magnetic exchange couplings as calculated by DFT on structures measured by XRD at 293 K and 3 K. (b) Pressure evolution of $J/J_{\varhexagon}$ and $J'/J_{\varhexagon}$ in the Y-kapellasite phase diagram for 293K and 3K.  Color gradient indicate the increase in pressure. Black square on the left represent DFT calculations from Ref. \cite{hering2022phase} on structure of Ref.~\cite{puphal2017strong}. Magenta star represents the couplings from INS at 1.5 K and ambient pressure in Ref.~\cite{chatterjee2023spin}. }
    \label{fig:jformulaDFT} 
\end{figure}

In the following we investigate the evolution of the  magnetic exchange couplings $J_i$ under hydrostatic pressure using DFT.
Calculations have been performed using crystal structures measured at six pressure values (0, 1.1, 3.0, 4.5, 5.9, 7.9 GPa) at room temperature (293 K), and on crystal structures measured at three different pressure values (0, 3.6, 7.0 GPa) at low temperature (3 K)~\cite{chatterjee2026emergence}.
The considered crystal structures  were refined with Olex2 \cite{olex} software, with the assumption of a constant O-H distance of 0.98 Å.
Prior calculations for this material~\cite{hering2022phase} demonstrated that only the first three nearest-neighbor (NN) exchange parameters are significant, while all others are negligible.
To verify this, we computed up to the $8^{\mathrm{th}}$ NN couplings and confirmed the persistence of this behavior under pressure (see Tables S7--S12 of the Supplemental Material).
Therefore, we restricted the calculations to these three exchange couplings. 
For consistency with previous literature, we henceforth denoted these as $J'$, $J_{\varhexagon}$, and $J$ respectively. The effective Heisenberg Hamiltonian thus reads
\[
\textit{H} = J'\sum_{\langle i j \rangle} \, \mathbf{S}_{i} \cdot \mathbf{S}_{j}
  + J_{\varhexagon}\sum_{\langle\!\langle i j \rangle\!\rangle} \, \mathbf{S}_{i} \cdot \mathbf{S}_{j}
  + J\sum_{\langle\!\langle\!\langle i j \rangle\!\rangle\!\rangle} \, \mathbf{S}_{i} \cdot \mathbf{S}_{j}\quad ,
\]
where ${\langle i j \rangle}$ for the first neighbors, $\langle\!\langle i j \rangle\!\rangle$ the second neighbors, and $\langle\!\langle\!\langle i j \rangle\!\rangle\!\rangle$ the third neighbors  [Fig.~\ref{fig:structure}(a)]. 
At 0 GPa, the experimentally known $\theta_{CW}$ of -100 K was obtained for a Hubbard  U = 9.27 eV for 293 K and U = 9.44 eV for 3 K (see Fig. S1 and Table S7 in Supplemental Material). Since the pressure dependence of the Curie temperature remains undetermined, this reference value of U was employed consistently to extract accurate exchange parameters across the studied pressure range.\\ Figure \ref{fig:jformulaDFT}(a) displays the pressure evolution of $J'$, $J_{\varhexagon}$, $J$, while Fig. \ref{fig:jformulaDFT}(b) shows how pressure affects $J'$/$J_{\varhexagon}$ and $J$/$J_{\varhexagon}$ on the phase diagram. Tables S4-S5 in Supplemental Material report the couplings values.
The crystal structures measured at 293 K and 3 K yield similar pressure dependences of the exchange couplings, with only slight differences in their magnitudes [Fig.~\ref{fig:jformulaDFT}(a)]. This behavior is consistent with the small structural changes between the two temperatures \cite{chatterjee2026emergence}.
Regarding pressure effects, only $J_{\varhexagon}$ is significantly affected, decreasing by about 20 K, while the other couplings remain nearly constant. This increases $J'/J_{\varhexagon}$ and $J/J_{\varhexagon}$, driving the system toward the spin liquid phase [Fig.~\ref{fig:jformulaDFT}(b)].

 \begin{figure}[]
    \centering
    \begin{overpic}[width=0.45\linewidth]{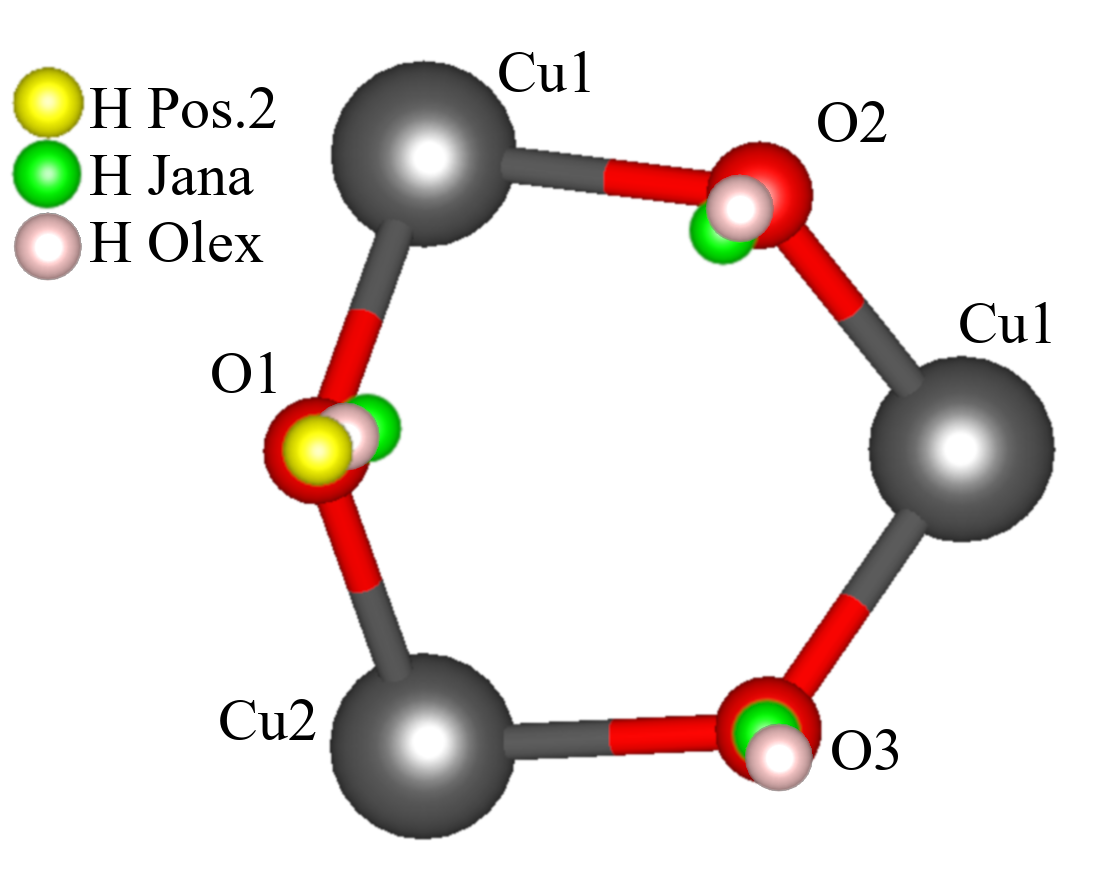}%
    \put(0,85){ \small (a)}  
    \end{overpic}
    \hfill
    \begin{overpic}[width=0.45\linewidth]{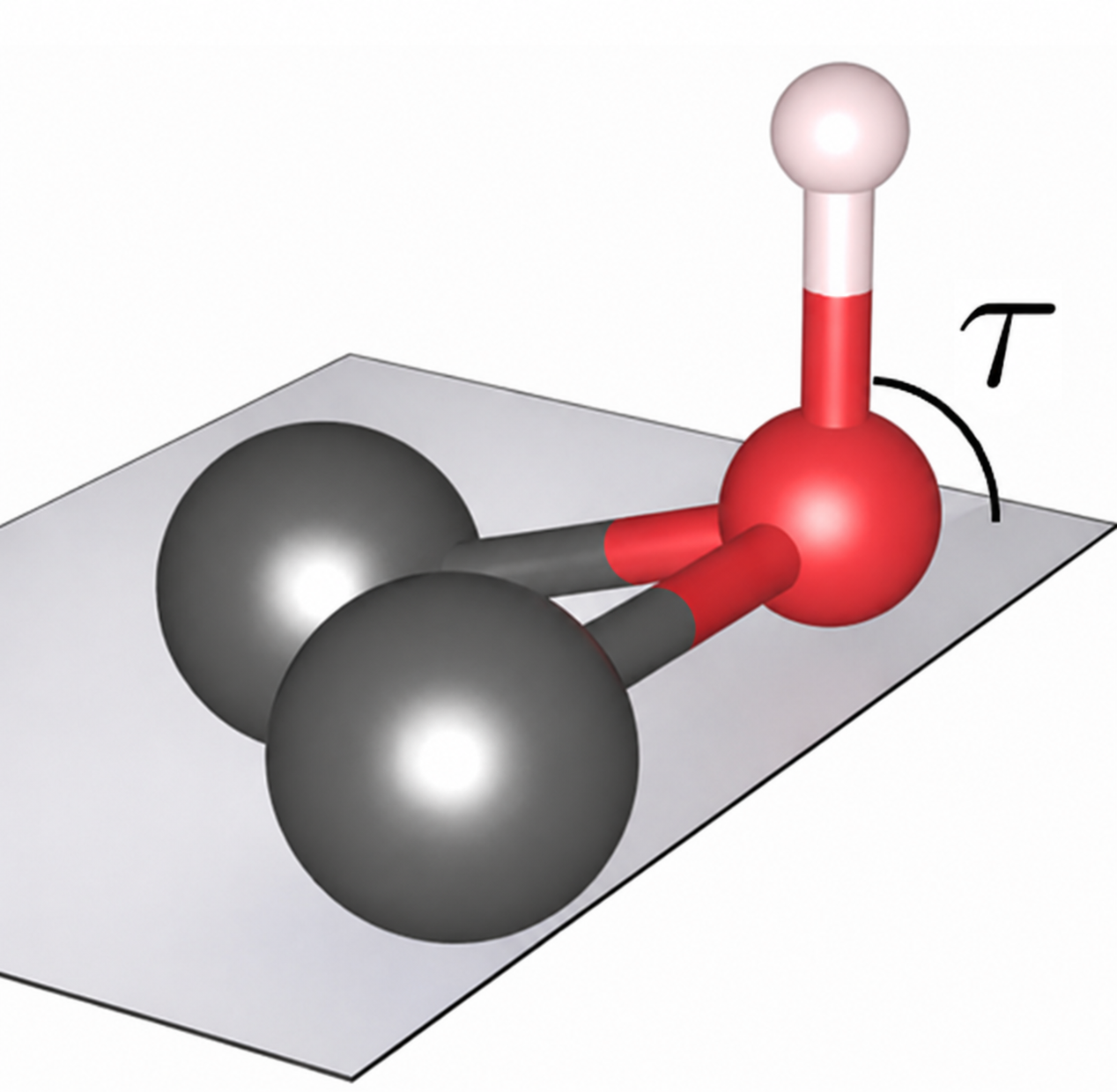}%
    \put(0,85){ \small (b)}    
    \end{overpic}
    \caption{(a) a-b plane view of the different hydrogen position considered in this study. The naming of the oxygens reflect the nearest-neighbor couplings, hence O1 is associated with $J'$, O2 with $J_{\varhexagon}$ and O3 with $J$ [see also Fig. \ref{fig:structure}]. In Pos.2 only H1 is manually moved, while the other H are left in the Olex2 position The different position are defined in terms of the out-of-plane angle $\tau$. (b) Visualization of out-of-plane angle $\tau$. The angle is defined as the complementary angle to that formed between the plane normal and the vector representing the O–H direction.}
    \label{fig:Hpos}
\end{figure}

\subsection{Influence of hydrogen position on the magnetic exchange couplings} 
\label{OHdist}
\begin{figure}[]
    \centering
    \begin{overpic}[width=0.99\linewidth]{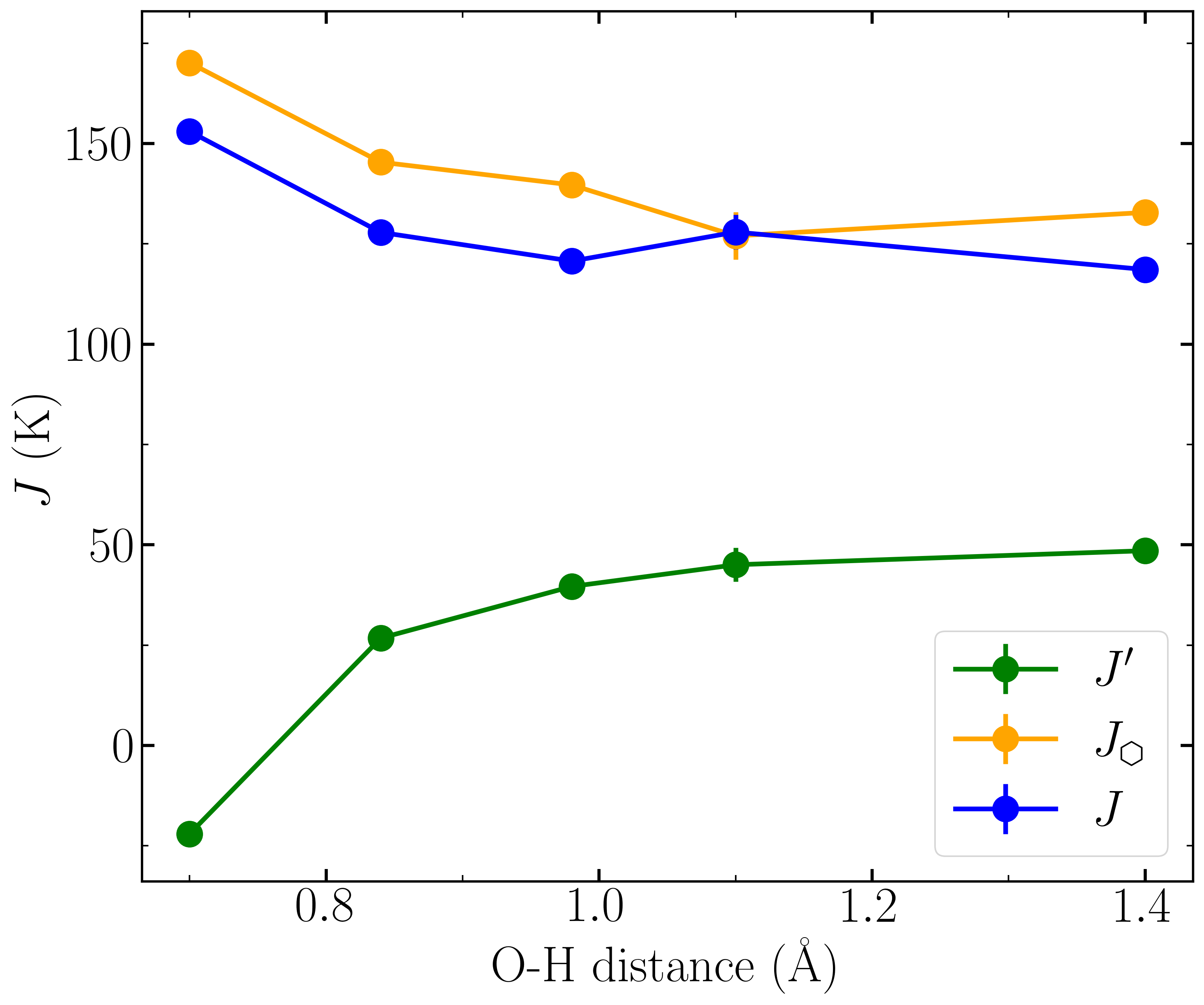}%
    \put(0,85){ \small (a)}
    \end{overpic}
    \par\vspace{0.5cm}
    \begin{overpic}[width=0.99\linewidth]{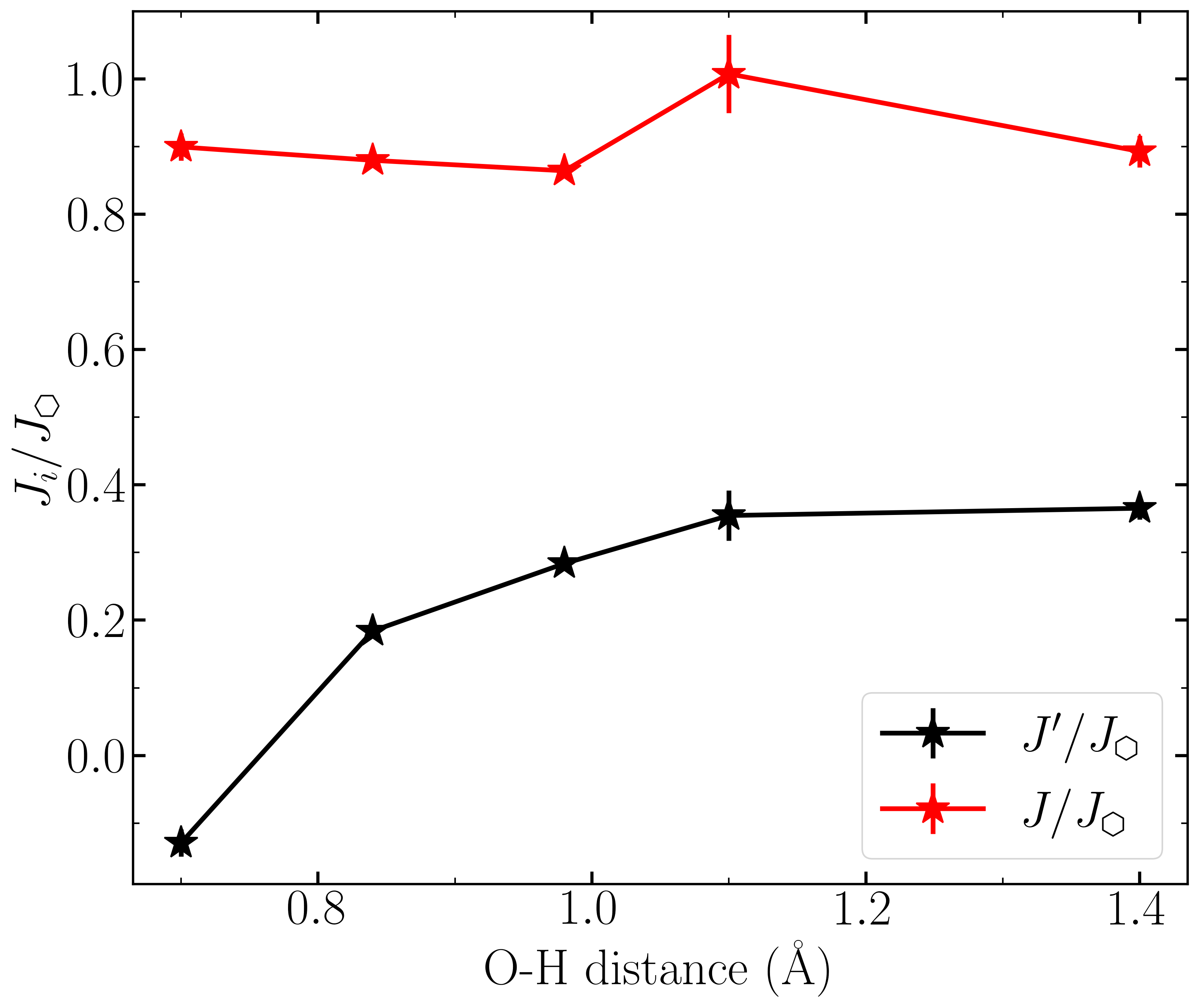}
    \put(0,85){ \small (b)}
    \end{overpic}
 
    \caption{(a) Exchange coupling behavior as a function of the O-H distance and (b) ratio of the exchange couplings as a function of the O-H distance, on the room temperature and ambient pressure structure. These ratios represent the x and y axis in the phase diagram. }
    \label{fig:Hdist}
\end{figure}

    This section investigates the effect of hydrogen positions on the calculated magnetic couplings. Because hydrogen scatters X-rays weakly, its position is difficult to determine experimentally and is often assigned using symmetry-based crystallographic procedures, which can yield different positions for the same structure [Fig.~\ref{fig:Hpos}(a)].  In particular the crystallographic software Olex2 uses the riding model, while Jana2020\cite{jana} performs a refinement based on geometrical considerations. \\
The hydrogen out-of-plane angle has already been shown to have an impact on the size of magnetic interaction on other systems \cite{lebernegg2011magneto}. 
To investigate the dependence of the exchange coupling on the O--H geometry, we performed a systematic scan of both the O--H bond length and its orientation, which is defined by the out-of-plane angle $\tau$, see Fig.~\ref{fig:Hpos}(b).
The effects of the out-of-plane angle are presented in Table ~\ref{tab:olexvsjana}, 
\begin{table}[t]%The best place to locate the table environment is directly after its first reference in text
\caption{\label{tab:olexvsjana}%
Magnetic exchange couplings for different out-of-plane angle $\tau$ as obtained with Olex2, Jana2020 and artificial Pos.2 crystal structure.  The structure have almost similar Cu-O-Cu angles but different $\tau$, showing the influence of this last parameter on the magnetic exchange couplings.}
\begin{ruledtabular}
\begin{tabular}{ccccccc}
\textrm{Pos}&
\textrm{$\tau_{J'}$ ($^\circ$)}&
\textrm{$\tau_{J\varhexagon}$ ($^\circ$)}&
\textrm{$\tau_{J}$ ($^\circ$)}&
\textrm{$J'$ (K)}&
\textrm{$J_{\varhexagon}$ (K)}&
\textrm{$J$ (K)}\\
\colrule
Jana &51.4& 55.6  &  53.5  &32.7(5)&110.6(4) &93.4(5) \\
Olex2 &46.9 & 45.2 &43.0 &41.4(1)&145.0(1) &125.6(1)  \\
Pos.2 &31.8 & 45.2 & 43.0 &84.8(7)&144.7(6) &114.4(7) \\
\end{tabular}
\end{ruledtabular}
\end{table}
where we calculated the magnetic couplings for the room-temperature (293 K) and ambient pressure (0 GPa) crystal structure refined by the two different softwares, and for an artificial position renamed Pos.2. Pos.2 was generated from the Olex2 structure by moving the hydrogen atoms H1 and aligning it with the underlying oxygen O1 in the $x$ and $y$ coordinates. A discussion about the crystal structure differences is presented in the Supplemental Material.  
The goal of studying this artificial position was to understand how the displacement of single hydrogen would affect the magnetic exchange. It has to be noted that H1 does not have any special property and the same study could be translated to the other H positions. In all cases, the structures yield almost identical lattice constants, Cu-Cu distance and Cu-O-Cu angle, to better disentangle hydrogen influence from other parameters.
A fixed on-site Hubbard interaction U = 9 eV was kept constant in this case, to highlight the changes induced by the new position and the O--H bond distance was fixed at 0.98 Å.\\
The results indicate that the exchange couplings decrease with increasing out-of-plane angle, while changes in a given hydrogen position mainly affect the corresponding coupling (e.g., $\tau_{J'}$ predominantly influences $J'$).\\
For the H--O bond length, we study the effects by manually changing the hydrogen atomic position while maintaining the same $\tau$ angle. In this case, we employed the room-temperature (293 K) and ambient-pressure (0 GPa) crystal structure refined by Olex2.
The results are presented in Fig.~\ref{fig:Hdist}, using the value of $U$ that reproduces the experimental $\theta_{CW}$. As shown in Fig.~\ref{fig:Hdist}(a), increasing the O--H distance enhances $J'$ while reducing $J_{\varhexagon}$ and $J$ by similar amounts. Consequently, $J/J_{\varhexagon}$ remains nearly constant, whereas $J'/J_{\varhexagon}$ increases and saturates around 0.4 [Fig.~\ref{fig:Hdist}(b)].

% \begin{figure}[]
%     \centering
%     \begin{overpic}[width=0.92\linewidth]{Figures/OH_J.png}
%      \put(0,78){ \small (a)} 
%     \end{overpic}
%     \par\vspace{0.3cm}
%     \begin{overpic}[width=0.92\linewidth]{Figures/OH_ratio.png}
%      \put(0,75){ \small (b)}   
%     \end{overpic}
%     \caption{(a) Exchange coupling behavior as a function of the O-H distance and (b) ratio of the exchange couplings as a function of the O-H distance, on the room temperature and ambient pressure structure. These ratios represent the x and y axis in the phase diagram. }
%     \label{fig:Hdist}
% \end{figure}

% \begin{table}[h]%The best place to locate the table environment is directly after its first reference in text
% \caption{\label{table:Hmov}%
% Magnetic exchange couplings for different hydrogen position at a fixed distance and fixed on-site Coulomb interaction U= 9 eV. Each position is identified by the out of plane angle $\tau$, with the subscript that states to which bond the hydrogen refers. The table shows how a small change in position of the hydrogen relative to $J'$ bond affects the magnitude of the interaction of the same bond ($J'$) and it is almost negligible with respect to the other bonds. }
% \begin{ruledtabular}
% \begin{tabular}{ccccc}
% \textrm{Position}&
% \textrm{$\tau_{J'}$ ($^\circ$)}&
% \textrm{$J'$ (K)}&
% \textrm{$J_{\varhexagon}$ (K)}&
% \textrm{$J$ (K)}\\
% \colrule
  
% Olex2 &46.9 &41.4(1.1)&145.0(1.1) &125.6(1.0)  \\
% POS.2 &31.8 &84.8(6.7)&144.7(6.3) &114.4(6.7) \\

% \end{tabular}
% \end{ruledtabular}
% \end{table}

\section{Discussion}

 As shown by the comparison between the calculations on structures obtained at 293 K versus 3 K in Fig.~\ref{fig:jformulaDFT}(a), temperature has only a minor effect on the magnetic exchange couplings. The largest differences occur at 0 GPa for $J'$ and $J_{\varhexagon}$, producing a small shift in the phase diagram that nearly vanishes under pressure. Given the larger errors on atomic positions of the low-temperature structure arising due to the employment of a cryostat, these differences may partly reflect experimental errors. Overall, the pressure dependence remains unchanged because the structural variations between 293 K and 3 K occur mainly along the \textit{c} axis, while the kagome \textit{ab} plane geometry is largely preserved. This allows low-temperature magnetic properties to be reliably inferred from room-temperature crystal structures.

Hydrostatic pressure primarily suppresses $J_{\varhexagon}$, whereas $J$ and $J'$ remain comparatively stable, leading to a crossover between $J_{\varhexagon}$ and $J$ [Fig.~\ref{fig:jformulaDFT}(a)]. This behavior closely follows the evolution of the corresponding Cu–O–Cu bond angles reported in Ref.~\cite{chatterjee2026emergence}, confirming that angle variations are the dominant factor controlling superexchange interactions. As a result, both $J'/J_{\varhexagon}$ and $J/J_{\varhexagon}$ increase with pressure, shifting Y-kapellasite towards the spin-liquid (SL) region of the phase diagram. These findings support hydrostatic pressure as an effective tuning parameter for approaching an SL state, although pressures above 10 GPa are likely required within the classical phase diagram. The results do not contradict the emergence of a fluctuating ground state at
lower pressure, because a full quantum mechanical treatment of the phase diagram, which potentially could shift the boundaries, has not yet been considered.

\begin{figure}[!ht]

    {\includegraphics[width=0.99\linewidth]{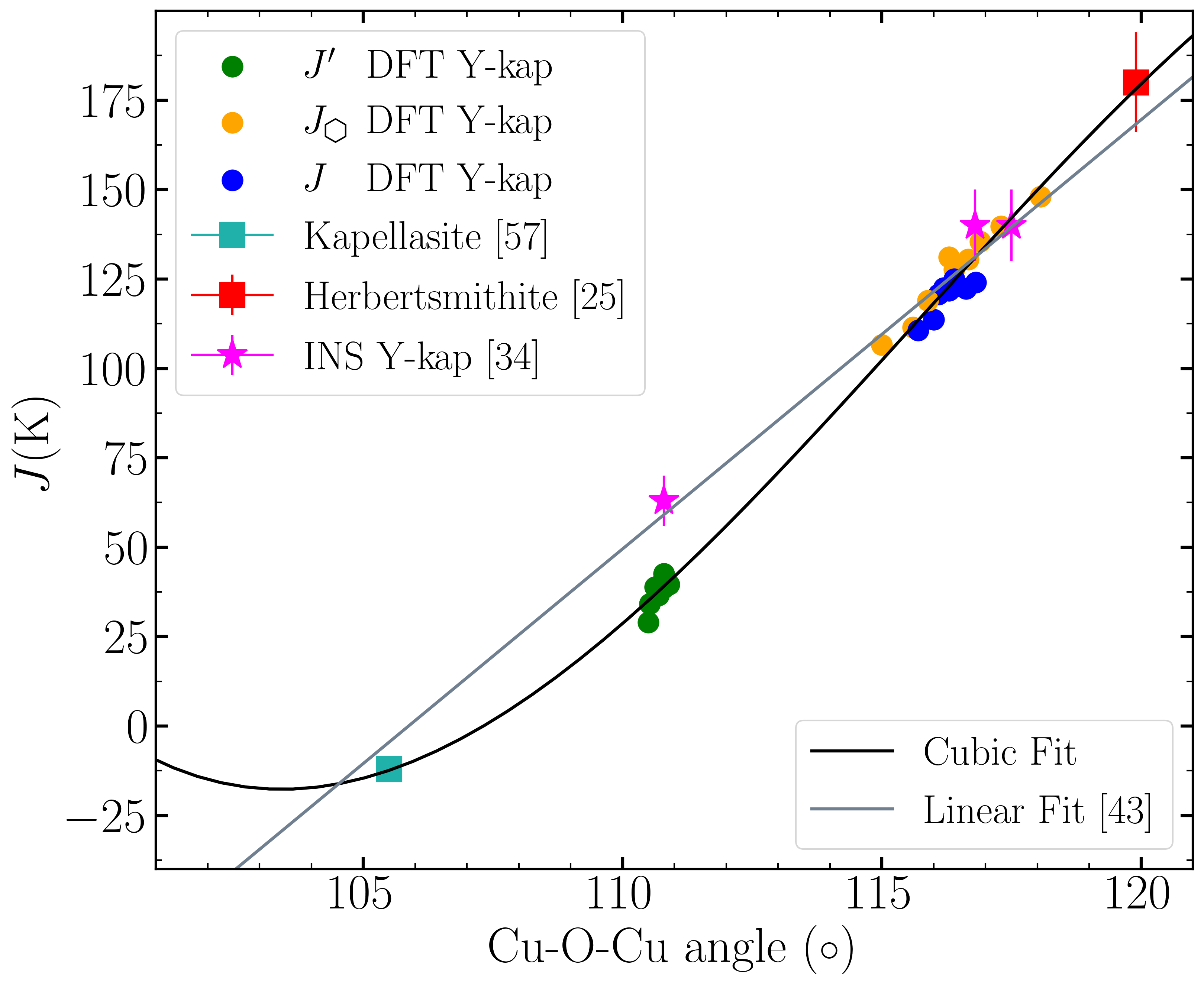}}%

    \caption{Exchange coupling vs Cu-O-Cu angle. The data points include the DFT calculated coupling of this manuscript, the three couplings extracted from INS \cite{chatterjee2023spin} and the result for Herbertsmithite \cite{khuntia2020gapless} and Kapellasite \cite{kermarrec2014spin}. INS $J'$ coupling (pink star around 110$^{\circ}$) is larger than the DFT calculated ones, while INS $J_{\varhexagon}$ and $J$ have perfect agreement. Gray line is the linear fit of Ref. \cite{chatterjee2026emergence} via the INS data points, while black line is the cubic fit proposed in this work based on the DFT values.}
    \label{fig:model}
\end{figure}

Compared with previous INS and DFT studies~\cite{chatterjee2023spin,hering2022phase}, our results predict a smaller $J'$, placing the system further from the SL boundary. While the analysis in Ref.~\cite{chatterjee2026emergence} employed a linear relationship between exchange couplings and Cu–O–Cu angles, our results are better described by a cubic dependence [Fig.~\ref{fig:model}]. This approach allows us to reproduce the behavior in the AFM regime \cite{rocquefelte2012theoretical,shimizuspin} and the characteristic features of the FM regime, appropriately weighted for Cu-hydroxide systems \cite{mizunoelectronic}. An accurate description on the FM regime it is beyond the scope of this manuscript. 
Despite the differences between the linear and cubic models, both approaches identify pressure-induced modifications of the Cu–O–Cu angle as the mechanism driving the system toward the spin-liquid (SL) regime.

The discrepancy with earlier DFT calculations can largely be attributed to the treatment of hydrogen positions. Our calculations confirm that increasing the hydrogen out-of-plane angle $\tau$ reduces the corresponding exchange coupling, in agreement with previous studies of hydroxy-bridged Cu dimers~\cite{lebernegg2011magneto}. The larger $\tau$ values adopted in Ref.~\cite{hering2022phase} naturally lead to smaller $J'$ and $J_{\varhexagon}$ and shift the system away from the SL boundary. In contrast, the $Olex2$ and $Jana2020$ refinements yield similar $\tau$ values and therefore nearly identical coupling ratios, despite differences in the absolute exchange strengths.

The O–H bond length mainly affects $J'/J_{\varhexagon}$, while $J/J_{\varhexagon}$ remains nearly unchanged. Even assuming a pressure-induced bond shortening of about 3\%, the resulting variation in $J'/J_{\varhexagon}$ is only ~4\%, validating the use of a fixed O–H distance in the refinements. Overall, these results highlight the crucial role of hydrogen geometry in determining exchange couplings in hydroxide-based superexchange systems~\cite{li2018role}, while confirming Cu–O–Cu angles as the primary parameter governing the pressure evolution towards the spin-liquid regime.

\section{Conclusion}

Using high-quality crystal structures measured under pressure at 3 K and 293 K~\cite{chatterjee2026emergence}, we performed \textit{ab initio} DFT calculations to investigate the evolution of magnetic exchange interactions in Y-kapellasite. Our results demonstrate that hydrostatic pressure efficiently tunes the magnetic Hamiltonian by modifying the Cu–O–Cu bond geometry, leading to a strong suppression of $J_{\varhexagon}$ and driving the system toward the spin-liquid region of the phase diagram. The pressure-induced evolution of the dominant exchange interaction follows a nonlinear dependence on the Cu–O–Cu bond angle, providing a microscopic explanation for the experimentally observed enhancement of magnetic frustration and suppression of long-range order. 

We further investigated the influence of hydrogen geometry and found that both the O–H bond length and the hydrogen out-of-plane angle $\tau$ significantly affect the magnetic exchange couplings. In particular, the sensitivity of $J'$ to hydrogen positioning largely accounts for discrepancies between previous DFT studies~\cite{hering2022phase} and highlights the importance of accurately determining hydrogen coordinates in hydroxide-based magnets. These findings establish hydrogen geometry as a key ingredient for reliable exchange-coupling calculations and for accurately assessing the proximity of Y-kapellasite to a quantum spin-liquid state. Future work combining crystallographic refinements beyond the independent atom model with neutron diffraction measurements could further improve hydrogen localization, which, together with an accurate determination of the phase boundaries within a full quantum mechanical picture, would provide a clearer picture of this compound.

\section{Acknowledgements}
AR and RV thank
the Deutsche Forschungsgemeinschaft (DFG, German
Research Foundation) for funding through TRR 288 —422213477 (Projects A05 and B05).
We acknowledge the ESRF within the in-house proposals BLC-15473, IH-HC-3746, IH-HC-4021 and IH-HC-4177.
We thank Apostolos Pantousas, Michael Merz and Max Gerin for the support and useful discussions.
\section{Data availability}
The DFT data is available at https://doi.org/10.5281/zenodo.21216749.
The high-pressure X-ray diffraction data from ID27 and ID15B is available at https://doi.esrf.fr/10.15151/ESRF-DC-2360737808.
The crystallographic structures on which the DFT was done are available from the joint CCDC/FIZ Karlsruhe database with deposition No. 2467984-2468000.
\bibliography{main.bib} 
\end{document}

\author{Federico Abbruciati}
 \email{federico.abbruciati@esrf.fr}
 \affiliation{European Synchrotron Radiation Facility, 71 Avenue des Martyrs, F-38043 Grenoble, France}
 \affiliation{Institute for Quantum Materials and Technologies, Karlsruhe Institute of Technology, Kaiserstr. 12, 76131
Karlsruhe, Germany.}

\author{Aleksandar Razpopov}
\email{razpopov@itp.uni-frankfurt.de}
\affiliation{Institut fur Theoretische Physik, Goethe-Universitat Frankfurt, 60438 Frankfurt am Main, Germany}

\author{João Elias F. S. Rodrigues}
\affiliation{European Synchrotron Radiation Facility, 71 Avenue des Martyrs, F-38043 Grenoble, France}

\author{Gaston Garbarino}
\affiliation{European Synchrotron Radiation Facility, 71 Avenue des Martyrs, F-38043 Grenoble, France}

\author{Matthieu Le Tacon}
\affiliation{Institute for Quantum Materials and Technologies, Karlsruhe Institute of Technology, Kaiserstr. 12, 76131
Karlsruhe, Germany.}

\author{Roser Valent\'\i}
\email{valenti@itp.uni-frankfurt.de}
\affiliation{Institut fur Theoretische Physik, Goethe-Universitat Frankfurt, 60438 Frankfurt am Main, Germany}

\author{Pascal Puphal}
\affiliation{Max Planck Institute for Solid State Research, Heisenbergstraße 1, D-70569 Stuttgart, Germany}
 
\author{Björn Wehinger}%
 \email{bjorn.wehinger@esrf.fr}
 \affiliation{European Synchrotron Radiation Facility, 71 Avenue des Martyrs, F-38043 Grenoble, France}

\date{\today}% It is always \today, today,
             %  but any date may be explicitly specified

\title{Supplemental material to: Ab Initio Investigation of Pressure Effects in the Spin-Liquid Candidate Y-Kapellasite }% Force line breaks with \\

\date{\today}% It is always \today, today,
             %  but any date may be explicitly specified
\maketitle

\section{Crystal structures}
\label{sec:single_crystal}
This section presents the structures refined with \textit{Olex2}\cite{olex} (already published in Ref. \cite{chatterjee2026emergence}) and \textit{Jana2020} \cite{jana}. Table \ref{table:olexvsjana} highlights the unit cell parameters and the relevant angle discussed in the main text, along with the value of the DFT calculated coupling.
Table \ref{table:struct1} and \ref{table:struct2} show the atomic position of the structures. 

\begin{table}[h!]%The best place to locate the table environment is directly after its first reference in text
\caption{\label{table:olexvsjana}%
Magnetic exchange couplings for \textit{Olex2} vs \textit{Jana2020} vs Pos.2 structures with fixed on-site Coulomb interaction U = 9 eV, as discussed in Sec. IIIB in the main text.  The structure have almost identical unit cell parameters and Cu-O-Cu $\phi$ angles but different $\tau$, showing the influence of this last parameter on the magnetic exchange couplings. Pos.2 was generated from Olex structure by moving the hydrogen related to $J'$.}

\begin{ruledtabular}
\begin{tabular}{cccccccccccc}
\textrm{Pos}&
\textrm{$a$ (Å)}&
\textrm{$c$ (Å)}&
\textrm{$\phi_{J'}$ ($^\circ$)}&
\textrm{$\phi_{J\hexagon}$ ($^\circ$)}&
\textrm{$\phi_{J}$ ($^\circ$)}&
\textrm{$\tau_{J'}$ ($^\circ$)}&
\textrm{$\tau_{J\hexagon}$ ($^\circ$)}&
\textrm{$\tau_{J}$ ($^\circ$)}&
\textrm{$J'$ (K)}&
\textrm{$J_{\hexagon}$ (K)}&
\textrm{$J$ (K)}\\
\colrule

Olex2 &11.57(0.1) &17.33(0.1)&110.9(2) & 117.3(2)&116.1(2)&46.9 & 45.2  & 43.0  & 41.4(1.1)&145.0(1.1) &125.6(1.0)  \\
Jana &11.55(0.1) &17.22(0.1)&110.9(2)  &117.8(2)&116.5(2)&51.4& 55.6  &  53.5 &32.7(4.6)&110.6(3.9) &93.4(5.1) \\
Pos.2&11.57(0.1) &17.33(0.1)&110.9(2) & 117.3(2)&116.1(2)& 31.8 & 45.2 & 43.0 &84.8(6.7)&144.7(6.3) &114.4(6.7) \\

\end{tabular}
\end{ruledtabular}
\end{table}

\begin{table*}[h]%The best place to locate the table environment is directly after its first reference in text
\caption{\label{table:struct1}%
Crystal structure parameters refined with \textit{Jana2020}\cite{jana} from X-ray single crystal diffraction data  measured at 293 K and 0 GPa. Spacegroup is R$\Bar{3}$ (No. 148) and unit cell parameters are $a$ = $b$ = 11.55(0.1) Å and $c$ = 17.22(0.1) Å.}
\begin{ruledtabular}
\begin{tabular}{lclll}
\textrm{}&
\textrm{Wick.}&
\textrm{x/$a$}&
\textrm{y/$b$}&
\textrm{z/$c$}\\
\colrule
Y1  & 6c  & 0.33333 & 0.66667 & 0.53830 \\
Y2  & 3b  & 0.00000 & 0.00000 & 0.50000 \\

Cu1 & 9d  & 0.50000 & 0.50000 & 0.50000 \\
Cu2 & 18f & 0.17468 & 0.33696 & 0.50346 \\

Cl1 & 6c  & 0.00000 & 0.00000 & 0.66164 \\
Cl2 & 18f & -0.00209 & 0.33565 & 0.61751 \\

O1  & 18f & 0.00780 & 0.19700 & 0.45650 \\
O2  & 18f & 0.15970 & 0.49100 & 0.46540 \\
O3  & 18f & 0.33890 & 0.47040 & 0.55700 \\
O4  & 3a  & 0.33333 & 0.66667 & 0.66667 \\

H1 & 18f & 0.01189 & 0.21813 & 0.40111 \\
H2 & 18f & 0.15936 & 0.49040 & 0.40847 \\
H3 & 18f & 0.34234 & 0.44364 & 0.61060 \\
H4 & 18f & 0.33420 & 0.59600 & 0.65600 \\

\end{tabular}
\end{ruledtabular}
\end{table*}
\begin{table*}[h]%The best place to locate the table environment is directly after its first reference in text
\caption{\label{table:struct2}%
Crystal structure parameters refined with \textit{Olex2} \cite{olex} from X-ray single crystal diffraction data measured at 293 K and 0 GPa. Spacegroup is R$\Bar{3}$ (No. 148) and unit cell parameters are $a$ = $b$ = 11.57(0.1) Å and $c$ = 17.23(0.1) Å.}
\begin{ruledtabular}
\begin{tabular}{lclll}
\textrm{}&
\textrm{Wick.}&
\textrm{x/$a$}&
\textrm{y/$b$}&
\textrm{z/$c$}\\
\colrule
Y1  & 6c  & 0.33333 & 0.66667 & 0.53830 \\
Y2  & 3b  & 0.00000 & 0.00000 & 0.50000 \\

Cu1 & 18f & 0.17468 & 0.33696 & 0.50346 \\
Cu2 & 9d  & 0.50000 & 0.50000 & 0.50000 \\

Cl1 & 6c  & 0.00000 & 0.00000 & 0.66164 \\
Cl2 & 18f & -0.00209 & 0.33565 & 0.61751 \\

O1  & 18f & 0.33890 & 0.47040 & 0.55700 \\
H1  & 18f & 0.34234 & 0.44364 & 0.61060 \\

O2  & 18f & 0.00780 & 0.19700 & 0.45650 \\
H2  & 18f & 0.01189 & 0.21813 & 0.40111 \\

O3  & 18f & 0.15970 & 0.49100 & 0.46540 \\
H3  & 18f & 0.15936 & 0.49040 & 0.40847 \\

O4  & 3a  & 0.33333 & 0.66667 & 0.66667 \\
H4  & 18f & 0.33420 & 0.59600 & 0.65600 \\

\end{tabular}
\end{ruledtabular}
\end{table*}

\section{Results of the main text}
Tables \ref{table:j_293K}-\ref{table:j_3K} report the evolution of the exchange couplings as a function of the pressure, as plotted in the main text in Fig. 2(a) and (b). Table \ref{table:Hdist} reports the values of the exchange couplings as a function of the O-H distance as reproduced in Fig. 3 of the main text.

\begin{table}[h]%The best place to locate the table environment is directly after its first reference in text
\caption{\label{table:j_293K}%
Pressure evolution of the magnetic exchange couplings calculated on the 293 K structure at a fixed U potential of 9.27 eV, which is the value that yields a $|\theta_{CW}|$ = 100 K on the ambient pressure structure.}
\begin{ruledtabular}
\begin{tabular}{ccccc}
\textrm{P (GPa)}&
\textrm{$J'$ (K)}&
\textrm{$J_{\hexagon}$ (K)}&
\textrm{$J$ (K)}&
\textrm{$|\theta_{CW}|$ (K)}\\
\colrule
0 &39.6(0.6)&139.7(0.7)&120.7(0.7)&100.0\\
1.1 &39.3(2.5)&135.5(1.7)&121.8(1.5)&98.9 \\
3.0 &38.7(0.7)&127.7(0.8)&122.5(0.8)&96.3\\
4.5 &38.9(0.7)&131.9(0.6)&125.4(0.5)&98.7\\
5.9 &40.9(1.4)&120.7(1.5)&118.0(1.4)&93.2\\
7.9&42.6(3.2)&106.6(1.5)&110.7(1.5)&86.6\\

\end{tabular}
\end{ruledtabular}
\end{table}

\begin{table}[h]%The best place to locate the table environment is directly after its first reference in text
\caption{\label{table:j_3K}%
Pressure evolution of the magnetic exchange couplings calculated on the 3 K structure at a fixed U potential of 9.44 eV, which is the value that yields a $|\theta_{CW}|$ = 100 K on the ambient pressure structure.}
\begin{ruledtabular}
\begin{tabular}{ccccc}
\textrm{P (GPa)}&
\textrm{$J'$ (K)}&
\textrm{$J_{\hexagon}$ (K)}&
\textrm{$J$ (K)}&
\textrm{$|\theta_{CW}|$ (K)}\\
\colrule
0 &29.0(5.0)&148.1(4.9)&122.8(5.0)&100.0\\
3.6 &34.2(0.7)&130.5(0.7)&122.3(0.7)&95.7\\
7.0 &38.9(3.0)&119.0(1.1)&124.1(3.0)&94.0\\

\end{tabular}
\end{ruledtabular}
\end{table}
\begin{table}[!]%The best place to locate the table environment is directly after its first reference in text
\caption{\label{table:Hdist}%
Magnetic exchange couplings calculated on the room temperature and ambient pressure structure, as a function of O-H distance at a fixed $|\theta_{CW}|$ Curie-Weiss temperature of 100 K. }

\begin{ruledtabular}
\begin{tabular}{ccccc}
\textrm{Distance (Å)}&
\textrm{$J'$ (K)}&
\textrm{$J_{\hexagon}$ (K)}&
\textrm{$J$ (K)}&
\textrm{$U$ (eV)}\\
\colrule
0.70 & -22.0(0.4) &170.1(0.5) &153.0(0.5)& 2.1 \\
0.84 &26.7(0.5) &145.3(0.6) & 127.9(0.5) & 6.6 \\
0.98 &39.6(0.6)&139.7(0.7)&120.7(0.7)    &9.3  \\
1.10 &45.0(4.2) &127.0(5.9) &127.9(4.3)  &  10.4\\
1.40&48.5(2.0) &132.8(2.7) &118.6(2.0)   &   13.6 \\

\end{tabular}
\end{ruledtabular}
\end{table}

\section{Full eight couplings}
Despite the couplings up to the $3^{rd}$ NN being the dominant ones \cite{hering2022phase} at ambient conditions, a sanity check is essential at higher pressure and low temperature to assess if this assumption is still valid.  Tables \ref{tab:eight293_0}-\ref{tab:three3_3.6} present the U dependence of the DFT calculated couplings on representative structures, up to the $8^{th}$ NN, proving the persistence of this behavior at the extreme conditions.

\begin{table}[H]%The best place to locate the table environment is directly after its first reference in text
\caption{Full eight coupling for 293 K at 0 GPa.
}
\label{tab:eight293_0}
\begin{ruledtabular}
\begin{tabular}{cccccccccc}
\textrm{U (eV)}&
\textrm{$J'$ (K)}&
\textrm{$J_{\hexagon}$ (K)}&
\textrm{$J$ (K)}&
\textrm{$J4$ (K)}&
\textrm{$J5$ (K)}&
\textrm{$J6$ (K)}&
\textrm{$J7$ (K)}&
\textrm{$J8$ (K)}&
\textrm{$|\theta_{CW}|$(K)}\\
\colrule

4 &72.2(1.4)&241.4(1.8)&230.6(1.8)&0.8(4.4)&2.9(2.0)&5.2(4.3)&2.9(2.0)&-2.7(6.5)&184.4\\
5 &66.0(1.2)&219.4(1.5)&206.5(1.5)&2.0(3.6)&2.9(1.6)&4.8(3.5)&2.9(1.6)&-1.3(5.3)&167.7\\
7 &53.9(0.9)&180.1(1.1)&163.7(1.1)&2.8(2.6)&3.9(2.5)&2.5(1.2)&3.9(2.5)&-0.1(3.9)&136.4\\
9 &41.6(0.7)&142.6(0.8)&124.3(0.8)&2.5(2.0)&1.8(0.9)&2.9(2.0)&1.8(0.9)&0.1(3.0)&105.8\\
10&35.3(0.6)&123.5(0.7)&105.4(0.7)&2.2(1.8)&1.4(0.8)&2.4(1.7)&1.4(0.8)& 0.1(2.6)&90.5\\	

\end{tabular}
\end{ruledtabular}
\end{table}

\begin{table}[H]%The best place to locate the table environment is directly after its first reference in text
\caption{Three coupling for 293 K at 0 GPa.
}
\label{tab:three293_0}
\begin{ruledtabular}
\begin{tabular}{ccccc}
\textrm{U (eV)}&
\textrm{$J'$ (K)}&
\textrm{$J_{\hexagon}$ (K)}&
\textrm{$J$ (K)}&
\textrm{$|\theta_{CW}|$(K)}\\
\colrule
4 & 71.4(2.0) & 246.0(2.1) & 232.9(2.3) & 183.4 \\
5 & 65.3(1.9) & 223.7(1.9) & 208.6(2.1) & 165.9 \\
7 & 53.4(1.5) & 183.6(1.6) & 165.4(1.7) & 134.1  \\
9 & 41.4(1.1) & 145.0(1.6) & 125.6(1.3) & 104.0 \\
10 & 35.2(0.9) & 125.4(0.9) & 106.4(1.0) & 89.0  \\

\end{tabular}
\end{ruledtabular}
\end{table}

\begin{table}[H]%The best place to locate the table environment is directly after its first reference in text
\caption{Full eight coupling for 293 K at 7.9 GPa.}
\label{tab:eight293_7p9}
\begin{ruledtabular}
\begin{tabular}{cccccccccc}
\textrm{U (eV)}&
\textrm{$J'$ (K)}&
\textrm{$J_{\hexagon}$ (K)}&
\textrm{$J$ (K)}&
\textrm{$J4$ (K)}&
\textrm{$J5$ (K)}&
\textrm{$J6$ (K)}&
\textrm{$J7$ (K)}&
\textrm{$J8$ (K)}&
\textrm{$|\theta_{CW}|$(K)}\\
\colrule

4&84.6(4.5) & 197.7(2.3) & 214.8(1.6) & 0.5(2.9)  & 13.3(7.78) &1.5(1.8) &  0.7(1.7)  & 1.5(1.8) & 171.5\\
5&75.3(4.7) & 177.4(2.3) & 192.2(1.7) & 0.1(3.0)  & 11.0(8.1)  &1.4(1.9) &  0.4(1.8)  & 1.4(1.9) & 153.0\\
7&59.9(5.5) & 142.4(2.7) & 152.7(1.9) & -0.9(3.5) & 8.7(9.4)   &0.8(2.2) &  -0.3(2.1) & 0.8(2.2) & 121.4\\
9&46.7(6.6) & 110.9(3.3) & 117.1(2.3) & -1.7(4.2) & 8.1(11.3)  &0.1(2.6) &  -1.3(2.5) & 0.1(2.6) & 93.3 \\

\end{tabular}
\end{ruledtabular}
\end{table}

\begin{table}[H]%The best place to locate the table environment is directly after its first reference in text
\caption{Three coupling for 293 K at  7.9 GPa.
}
\label{tab:three293_7p9}
\begin{ruledtabular}
\begin{tabular}{c c c c c}
\textrm{U (eV)}&
\textrm{$J'$ (K)}&
\textrm{$J_{\hexagon}$ (K)}&
\textrm{$J$ (K)}&
\textrm{$|\theta_{CW}|$(K)}\\
\colrule
4&83.1(3.1) & 200.1(1.5) & 216.0(1.4)&166.4\\
5&73.6(2.9) & 179.4(1.4) & 193.1(1.3)&148.7\\
7&58.5(3.0) & 144.3(1.4) & 153.4(1.3)&118.7\\
9&45.9(3.7) & 113.2(1.7) & 118.0(1.6)&92.4\\

\end{tabular}
\end{ruledtabular}
\end{table}

\begin{table}[H]%The best place to locate the table environment is directly after its first reference in text
\caption{Full eight coupling for 3 K at 3.6 GPa.
}
\label{tab:eight3_3.6}
\begin{ruledtabular}
\begin{tabular}{cccccccccc}
\textrm{U (eV)}&
\textrm{$J'$ (K)}&
\textrm{$J_{\hexagon}$ (K)}&
\textrm{$J$ (K)}&
\textrm{$J4$ (K)}&
\textrm{$J5$ (K)}&
\textrm{$J6$ (K)}&
\textrm{$J7$ (K)}&
\textrm{$J8$ (K)}&
\textrm{$|\theta_{CW}|$(K)}\\
\colrule
4	&59.7(3.5)&245.8(4.1)&251.5(3.5)&3.9(7.3)&5.7(9.8)&2.2(4.5)&11.6(9.7)&2.2(4.5)&194.2\\
5	&59.0(0.4)&222.0(0.5)&220.6(0.4)&4.2(0.8)&3.8(1.1)&2.1(0.5)&1.8(1.1) &2.1(0.5)&171.9\\
7	&47.4(0.3)&179.6(0.4)&174.6(0.3)&3.1(0.6)&4.5(0.8)&1.6(0.4)&1.1(0.8) &1.6(0.4)&137.8\\
9	&36.6(0.3)&140.7(0.4)&133.3(0.3)&2.1(0.7)&4.7(0.9)&1.1(0.4)&0.1(0.9) &1.1(0.4)&106.6\\
\end{tabular}
\end{ruledtabular}
\end{table}

\begin{table}[H]%The best place to locate the table environment is directly after its first reference in text
\caption{Three coupling for 3 K at 3.6 GPa.
}
\label{tab:three3_3.6}
\begin{ruledtabular}
\begin{tabular}{ccccc}
\textrm{U (eV)}&
\textrm{$J'$ (K)}&
\textrm{$J_{\hexagon}$ (K)}&
\textrm{$J$ (K)}&
\textrm{$|\theta_{CW}|$(K)}\\
\colrule
4   &60.7(3.5)   &250.6(3.4)	&252.5(3.5)	&187.9\\
5   &59.8(1.0)   &223.8(1.0)	&221.4(1.0)	&168.3\\
7   &48.0(0.8)   &80.7(0.8)	    &175.2(0.8)	&134.6\\
9   &37.0(0.6)   &41.2(0.6)	    &133.8(0.6)	&104.0\\
			
\end{tabular}
\end{ruledtabular}
\end{table}

\section{U dependence} 
\begin{figure*}[b]
    \begin{overpic}[scale=0.25]{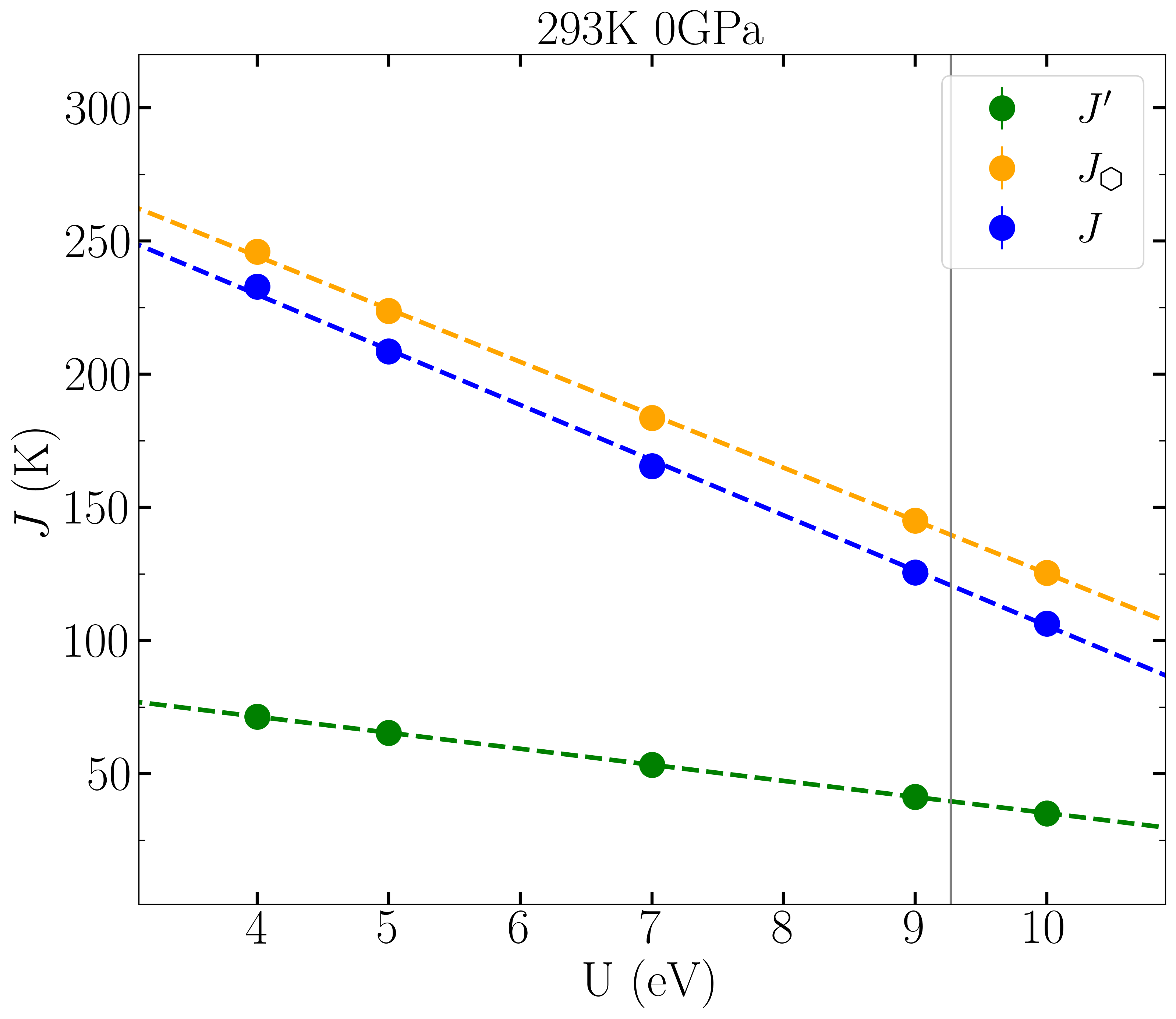}%
    \put(0,85){ \small (a)}
    \end{overpic}
    \begin{overpic}[scale=0.25]{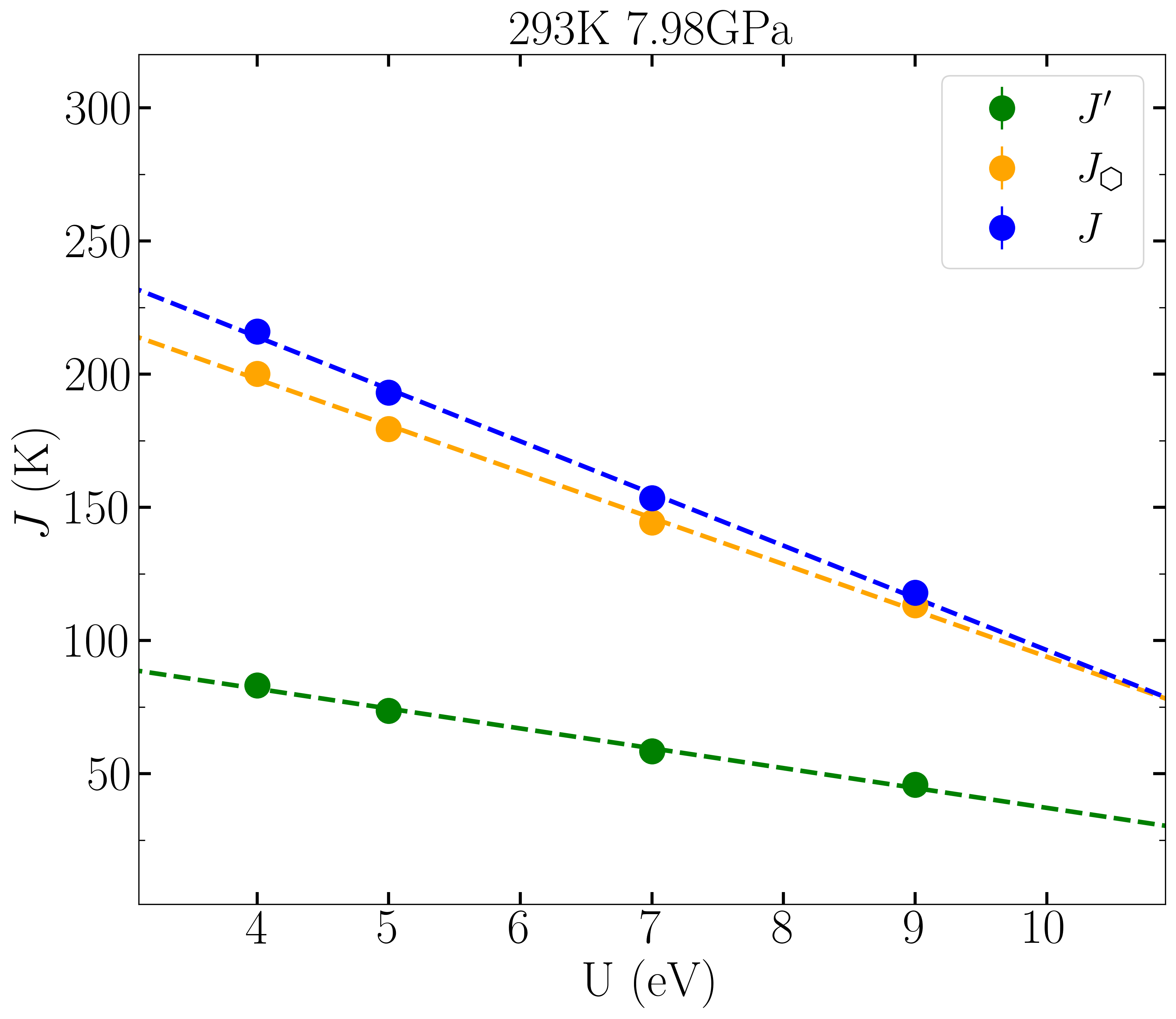}%
    \put(0,85){ \small (b)}
    \end{overpic}
    \begin{overpic}[scale=0.25]{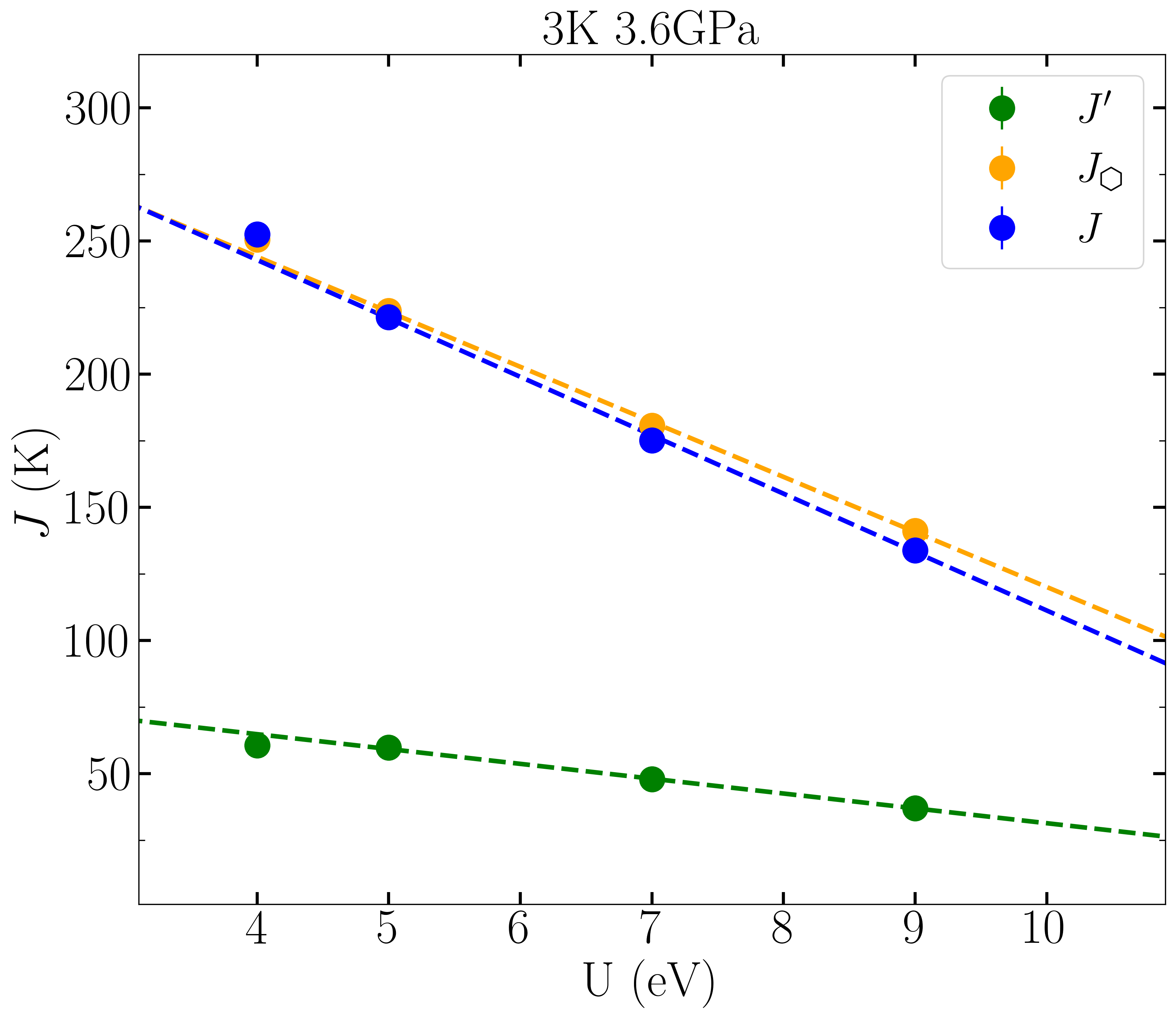}%
    \put(0,82){ \small (c)}
    \end{overpic}

    \caption{U dependence of the magnetic couplings for the selected structures presented in the section before. Vertical line in (a) highlights the U potential that yields a $\theta_{CW}$ of -100 K, which is U = 9.27 eV.}
    \label{figure:fit293K}
\end{figure*}

\begin{figure*}[b]
    \begin{overpic}[scale=0.25]{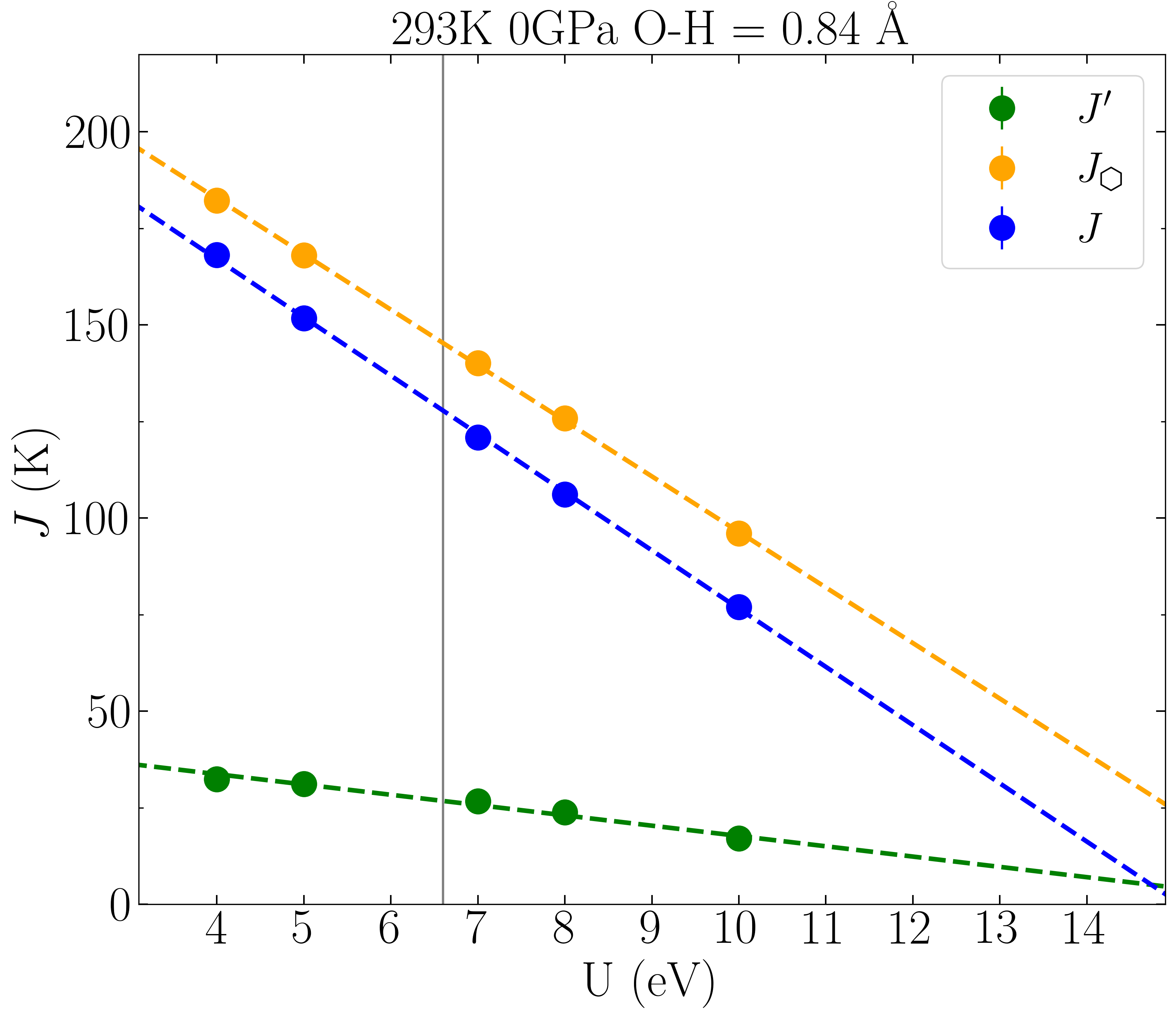}%
    \put(0,85){ \small (a)}
    \end{overpic}
    \begin{overpic}[scale=0.25]{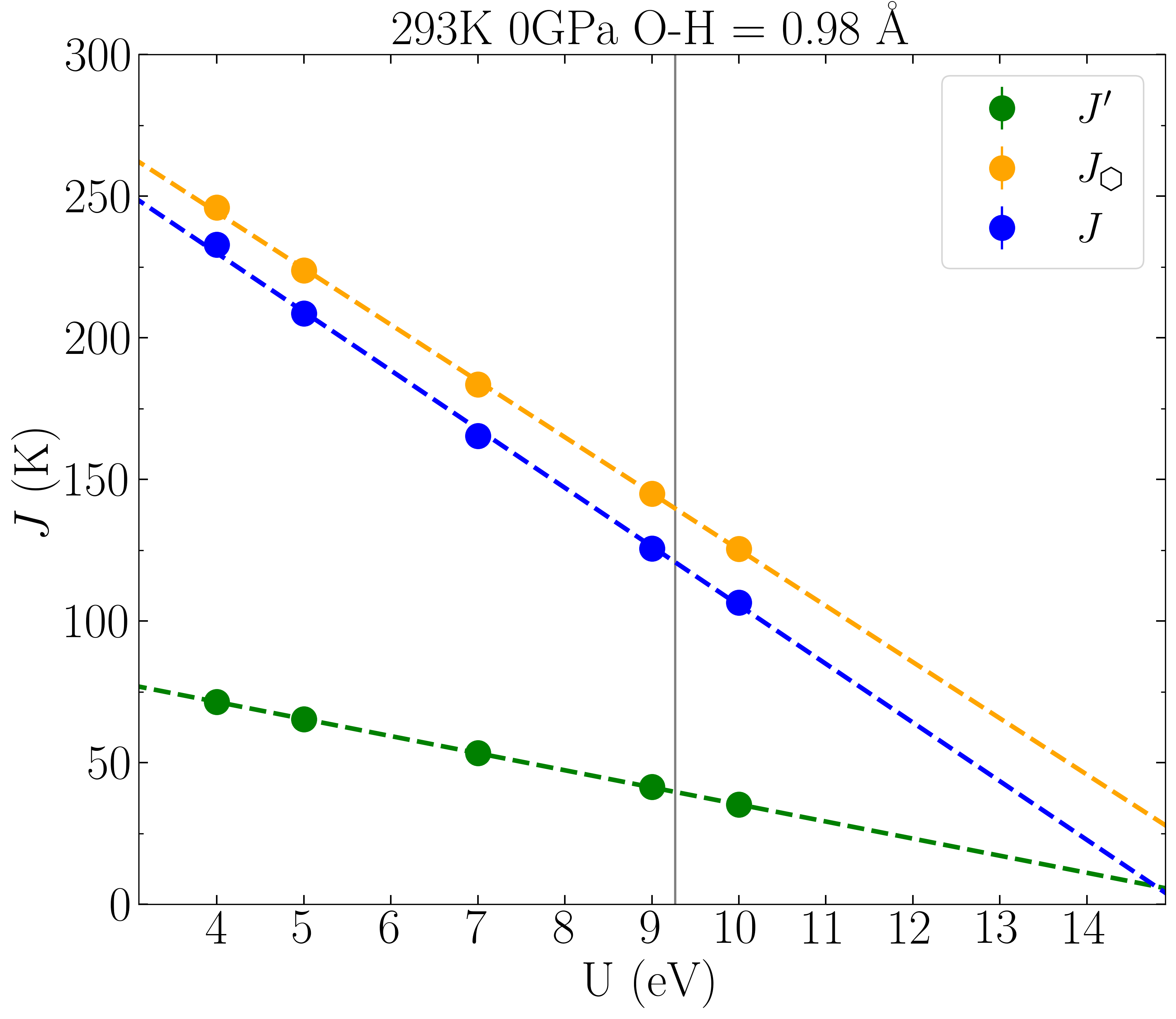}%
    \put(0,85){ \small (b)}
    \end{overpic}
    \begin{overpic}[scale=0.25]{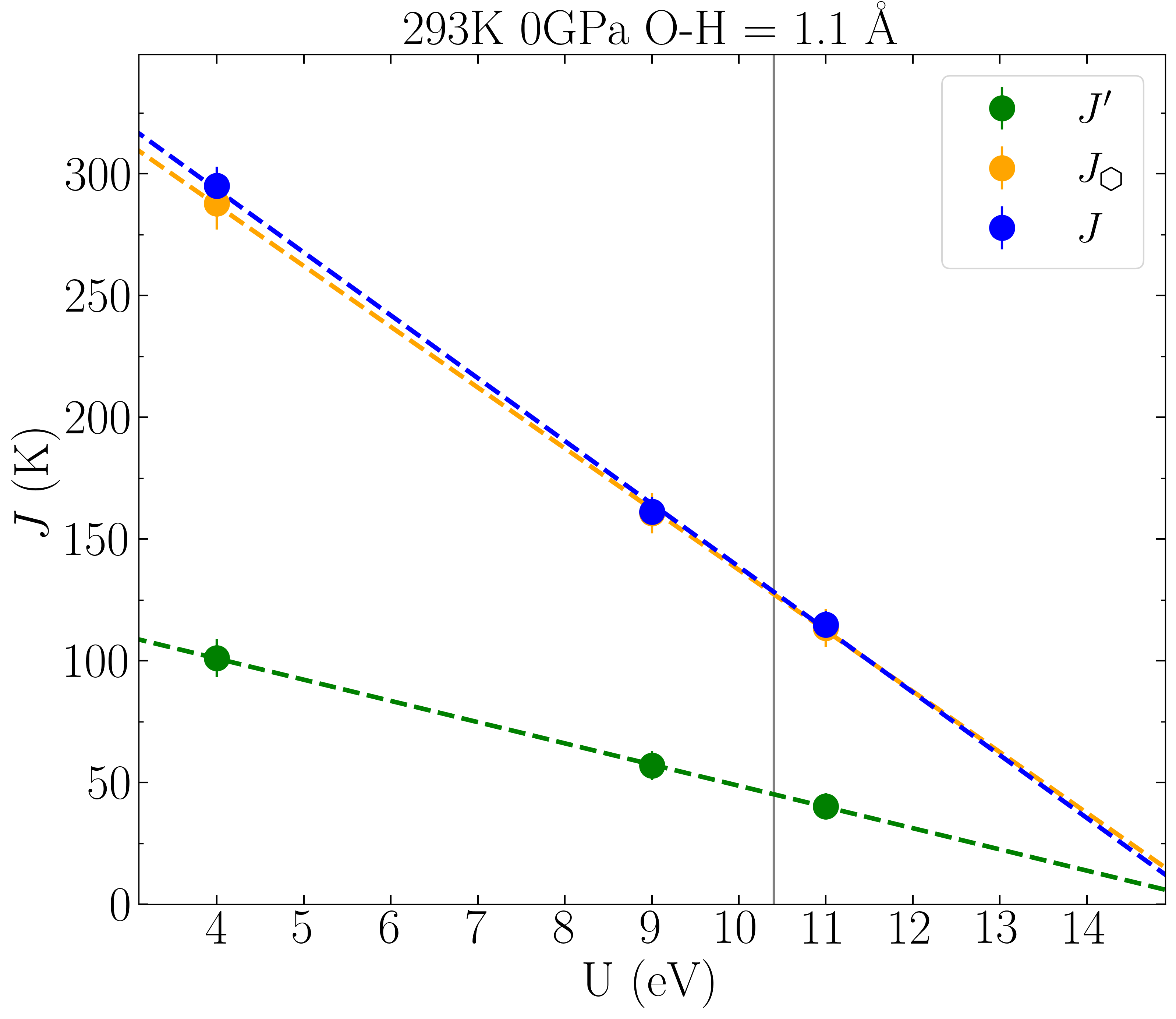}%
    \put(0,82){ \small (c)}
    \end{overpic}

    \caption{U dependence of the magnetic couplings for each O-H distance. Vertical line in highlights the U potential that yields a $|\theta_{CW}|$ of 100 K.}
    \label{figure:fit5}
\end{figure*}

This section goes through the U dependence of the magnetic couplings. For each structure, the couplings were calculated for a set of discrete U values [see Tables \ref{tab:eight293_0}-\ref{tab:three3_3.6}]. The dependence of the couplings to U was found by a linear fitting, as shown in Fig.\ref{figure:fit293K}. The exchange coupling values reproducing the experimental $\theta_{CW}$ were then extracted. The relationship between $\theta_{CW}$ and $J$ is evaluated in a mean field approach \cite{suzukimean} following the formula
\[
\theta_{\mathrm{CW}}(K) =- \frac{4}{9} s(s+1)\sum_i z_i J_i =- \frac{1}{3}\sum_i J_i.
\]
which is extracted from a mean-field framework. z is the number of nearest-neighbor spins. The $\theta_{CW}$ under pressure is not experimentally known. Therefore we employed a constant U potential, which was the one that reproduced $\theta_{CW}$ = -100 K at ambient pressure. 
Instead, for the O-H distance analysis, the couplings were extracted at the experimentally known  $\theta_{CW}$  as shown in Fig. \ref{figure:fit5}.

%  \begin{figure*}[h!]
%     \centering
%     {\includegraphics[scale=0.13]{Supp_figures/fit 3_0_label.png}}%
%     \quad
%     {\includegraphics[scale=0.13]{Supp_figures/fit 3_3_label.png}}%
%     \vspace{0.5cm}
%     {\includegraphics[scale=0.13]{Supp_figures/fit 3_7_label.png}}%

%     \caption{U dependence of the magnetic couplings for each pressure at 3 K. Vertical line in (a)  highlights the U potential that yields a $|\theta_{CW}|$ of 100 K, which is U = 9.44 eV. Since the pressure dependence of $\theta_{CW}$ is not experimentally known, once the correct value of U was found at ambient pressure, it was kept constant for the extraction of the coupling at all other pressure.}
%     \label{figure:fit3K}
% \end{figure*}

\clearpage
\bibliography{supp}